# Dual slow-light enhanced photothermal gas spectroscopy on a silicon chip

Kaiyuan Zheng,[1,2,3,†,#] Zihang Peng,[1,2,3,†] Hanyu Liao,[2,3,†] Yijun Huang,[1] Haihong Bao,[2,3] Shuangxiang Zhao,[2,3] Yu Zhang,[1] Chuantao Zheng,[1,*] Yiding Wang,[1] Wei Jin[2,3,*]

[1]State Key Laboratory of Integrated Optoelectronics, College of Electronic Science and Engineering, Jilin University, 2699 Qianjin Street, Changchun, 130012, China
[2]Department of Electrical and Electronic Engineering and Photonics Research Institute, The Hong Kong Polytechnic University, Hung Hom, Hong Kong, 999077, China
[3]Photonics Research Center, The Hong Kong Polytechnic University Shenzhen Research Institute, 18 Yuexing Road, Shenzhen, 518071, China
[†]These authors contributed equally to this work
[#]Present address: Division of Environment and Sustainability, The Hong Kong University of Science and Technology, Clear Water Bay, Hong Kong, 999077, China
[*]Corresponding authors: zhengchuantao@jlu.edu.cn, wei.jin@polyu.edu.hk

**Abstract:** Integrated photonic sensors have attracted significant attention recently for their potential for high-density integration. However, they face challenges in sensing gases with high sensitivity due to weak light-gas interaction. Slow light, which dramatically intensifies light-matter interaction through spatial compression of optical energy, provides a promising solution. Herein, we demonstrate a dual slow-light scheme for enhancing the sensitivity of photothermal spectroscopy (PTS) with a suspended photonic crystal waveguide (PhCW) on a CMOS-compatible silicon platform. By tailoring the dispersion of the PhCW to generate structural slow light to enhance pump absorption and probe phase modulation, we achieve a photothermal efficiency of $3.6\times10^{-4}$ rad·cm·ppm$^{-1}$·mW$^{-1}$·m$^{-1}$, over 1−3 orders of magnitude higher than the strip waveguides and optical fibers. With a 1-mm-long sensing PhCW incorporated in a stabilized on-chip Mach-Zehnder interferometer with a footprint of 0.6 mm$^2$, we demonstrate acetylene detection with a sensitivity of $1.4\times10^{-6}$ in terms of noise-equivalent absorption and length product (NEA·L), the best among the reported photonic waveguide gas sensors to our knowledge. The dual slow-light enhanced PTS paves the way for integrated photonic gas sensors with high sensitivity, miniaturization, and cost-effective mass production.

Integrated photonics have the potential for widespread applications in communications, sensing, metrology, and information processing.[1,2] Silicon-based on-chip gas sensing is an emerging field of integrated photonics, which has attracted increasing attention due to its unique advantages in size, weight, power efficiency and cost, but faces challenge in detection sensitivity due to inherently weak light-matter interaction.[3,4] Slow light offers an effective means for improving light-matter interaction on the photonic chip, which arises from strong interference between the forward-propagating and back-scattered waves in a photonic crystal,[5-8] leading to enhanced electromagnetic field intensity.[9-12]

    Photothermal spectroscopy (PTS) is a sensitive gas detection method. It detects the phase change of a probe beam due to the reflective index (RI) change of the gas medium, originating from the gas absorption of a modulated pump beam.[13-15] The sensitivity performance of PTS gas sensors may be quantified by defining the normalized photothermal (PT) efficiency in the unit of rad·cm·ppm$^{-1}$·mW$^{-1}$·m$^{-1}$, indicating the magnitude of probe phase modulation per unit gas

concentration (ppm), absorption coefficient (cm$^{-1}$), absorption distance (m), and pump power (mW).[16,17] Early works on free-space-based PTS have demonstrated PT efficiency of ~$10^{-10}$ rad·cm·ppm$^{-1}$·mW$^{-1}$·m$^{-1}$ level.[18,19] Optical fibers and waveguides provide much tighter light confinement than free space, which enhance the optical field intensity for light-gas interaction and hence the PT efficiency.[20] Silica hollow-core fibers (HCFs) with a core diameter of a few tens of micrometers have demonstrated PT efficiency of ~$10^{-6}$ rad·cm·ppm$^{-1}$·mW$^{-1}$·m$^{-1}$,[21,22] and an on-chip chalcogenide waveguide with core-thickness of ~200 nm demonstrated PT efficiency of ~$10^{-5}$ rad·cm·ppm$^{-1}$·mW$^{-1}$·m$^{-1}$.[23]

In this work, we propose a dual slow-light enhanced PTS on a suspended silicon photonic crystal waveguide (PhCW) to dramatically improve the on-chip PT efficiency. We leverage the PhCW to generate strong slow-light effect by tailoring the structural dispersion.[24] The dual slow-light region is designed to cover both the pump and probe wavelengths, where the pump slow light enhances the optical field intensity for gas absorption, and the probe slow light amplifies the PT-induced phase modulation. A suspended structure is designed to achieve higher light-gas overlap as well as better heat accumulation by using low-thermal-conductivity air as the top and bottom claddings. A silicon-integrated PTS gas sensor with dual slow-light enhancement demonstrates on-chip PT efficiency of $3.6 \times 10^{-4}$ rad·cm·ppm$^{-1}$·mW$^{-1}$·m$^{-1}$, over 1−3 orders of magnitude higher than the strip waveguides and optical fibers. The PT efficiency is proportional to the product of group indices of the pump and probe, which can be further enhanced by dispersion optimization.

## Results

**Theory.** Fig. 1(a) illustrates the basics of the dual slow-light enhanced PTS on a suspended silicon two-dimensional (2D)-PhCW. The PhCW is composed of air holes etched into hexagonal lattices of silicon, and the suspended structure is formed by removing the silica bottom cladding layer. The slow-light modes, generated by a line defect, propagate along the $z$-direction. These modes are confined in the silicon layer by total internal reflection in the $x$-direction and localized in the defect by photonic bandgap in the $y$-direction. A pump and a probe slow-light modes co-propagate along the PhCW, where the modulated pump absorption by gas molecules induces a heat source that changes the temperature distribution and hence modulates the phase of the probe beam. The optical intensities of both beams are compressed spatially by the slow-light effect in the PhCW, leading to enhanced efficiency of the pump absorption and probe phase modulation. The heat-source ($\tilde{Q}$) over the waveguide cross-section ($x$-$y$) at propagation distance $z$ equals to the absorbed pump intensity through Beer-Lambert law and is given by

$$\tilde{Q}(x,y) = \tilde{I}^{(0)}\left[1 - \exp\left(-\frac{n_{gp}}{n_{gas}^{(0)}}\alpha C \gamma z\right)\right] \propto \frac{n_{gp}}{n_{gas}^{(0)}} \alpha C \gamma P_p \quad (1)$$

where $n_{gas}^{(0)}$ and $\tilde{I}^{(0)}$ are the gas RI and modulated optical intensity of the pump beam with superscript (0) denotes the values in the absence of gas absorption, $\alpha$ the gas absorption coefficient, $C$ the gas concentration. The "~" denotes the frequency-domain amplitude of the modulated variables. $n_g$ is the group index directly related to the energy transport velocity, i.e., the group velocity in the absence of loss.[25] $n_{gp}$ represents the group index at pump wavelength $\lambda_p$, which enhances $\tilde{Q}$ via the slow-light effect. As shown in Fig. 1(b), the heat source also depends on the fractional power of pump mode in gas, quantified as $\gamma P_p$. Here the spatial confinement factor $\gamma$ represents the fraction of energy that interacts with the surrounding gas (~13%),[26] and $P_p$ is the incident pump power.

The heat source alters the temperature distribution ($\tilde{T}$) through heat conduction. For an intensity-modulated pump beam, the amplitude of the temperature oscillation may be approximated as $\tilde{T} \propto \tilde{Q}/\kappa$, with $\kappa$ the thermal conductivity of the waveguide.[20] The profile of the temperature change is shown in Fig. 1(c). Due to the low thermal conductivity of the gas medium (~0.02 W/m/K), the suspended structure lowers the heat conduction in the *x*-direction and facilitates the heat accumulation. Due to significant overlap with the probe field, as shown in Fig. 1(d), the modulated $\tilde{T}$ changes the phase of the probe beam via the PT-induced RI variation. The slow-light enhanced PT phase modulation amplitude ($\Delta\tilde{\phi}$) is approximated as:

$$\Delta\tilde{\phi} \approx \frac{2\pi n_{gb} L}{\lambda_b n_{Si}} e_{Si}^{TO} \langle \psi_b | \tilde{T} | \psi_b \rangle_{Si} \tag{2}$$

where $n_{gb}$ is the group index at the probe wavelength $\lambda_b$, $L$ the PhCW length, $\psi_b$ the normalized probe mode distribution, $e_{Si}^{TO}$ the thermo-optic (TO) coefficient of silicon (~1.8×10$^{-4}$ K$^{-1}$). We neglect the contribution of gas-induced TO effect due to its very small TO coefficient (−0.91×10$^{-6}$ K$^{-1}$) as compared to that of silicon with relative error <1% (Fig. S1). The probe phase modulation is primarily decided by the spatial integration of $\psi_b$ and $\tilde{T}$ in the silicon region. From Eqs. (1) and (2), the phase modulation is related to the product of $n_{gp}$ and $n_{gb}$, given by:

$$\Delta\tilde{\phi} = k^* \alpha C L P_p \propto \frac{2\pi n_{gp} n_{gb} L}{\lambda_b \kappa_{Si}} e_{Si}^{TO} \alpha C \gamma P_p \tag{3}$$

here $k^*$ is defined as the normalized PT efficiency in the unit of rad·cm·ppm$^{-1}$·mW$^{-1}$·m$^{-1}$, representing the phase modulation (in rad) of the probe beam normalized by $\alpha C P_p L$ (in cm$^{-1}$·ppm·mW·m). As shown in Eq. (3), the slow-light effect of PhCW greatly improves the PT phase modulation, which is attributed to the enhanced $n_{gp}$ that amplifies the heat generation and $n_{gb}$ that amplifies the probe phase modulation. Detailed theory refers to Supplementary S1.

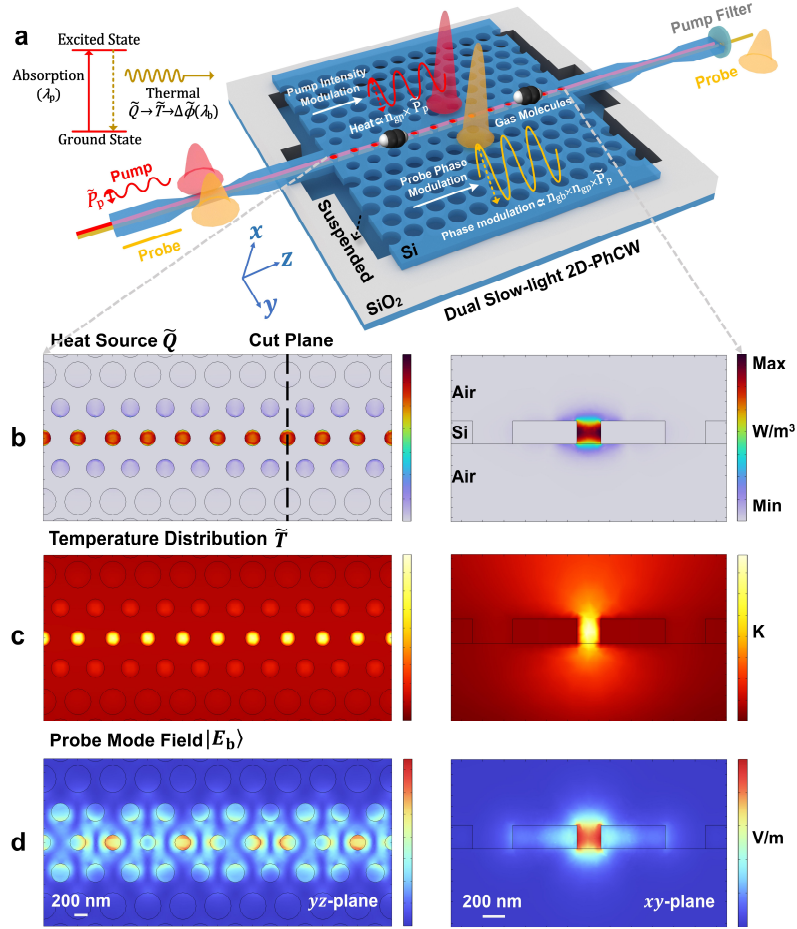

**Fig. 1 Principle of the dual slow-light enhanced PTS. (a)** Schematic of dual slow-light enhancement on a suspended 2D-PhCW. Pump and probe beams are launched into the PhCW with their energy compressed by the slow-light effect. Gas molecules absorb the compressed pump energy and are promoted to an excited energy level, which subsequently return to the ground state via thermal relaxation. This acts as **(b)** a heat source $\tilde{Q}$, which induces **(c)** a temperature change $\tilde{T}$ on surrounding materials through thermal conduction. **(d)** A probe mode field $|E_b\rangle$ is used to probe the temperature change as a phase modulation, which depends on the overlap of $\tilde{T}$ with $|E_b\rangle$ as well as the slow-light enhancement at the probe wavelength. This dual slow-light effect results in an overall enhancement of $n_{gp} \times n_{gb}$ on the probe phase modulation $\Delta\tilde{\phi}$. The suspended design of the PhCW enhances the light-gas interaction and the heat accumulation. For simplicity, the variables shown in (b-d) are normalized against their maxima.

**Simulation of PhCW PTS.** The PhCW is on a silicon-on-insulator (SOI) wafer with a 220-nm-thick top-silicon layer and three rows of defect holes. The lattice constant ($a$) and radius of the defect holes ($R$, $r_0$, $r_1$) are optimized using the RSoft band-solver (Supplementary S2). The dispersion diagram of the PhCW is depicted in Fig. 2(a) with optimal parameters of $a$ = 500 nm, $R$ = 0.38$a$, $r_0$ = 0.6$R$ and $r_1$ = 0.7$R$. There is an even mode and an odd mode in the photonic bandgap. The even mode, easily excited by the fundamental TE mode of the input strip waveguide, is used for gas sensing. The slow-light wavelength band includes pump and probe regions. The pump region covers the $C_2H_2$ vibrational-rotational absorption band (1515−1543 nm), and the probe region is in the non-absorption band (1543−1572 nm). The group index of the slow-light mode can be expressed as $n_g = c \cdot d\mathbf{k}/d\omega$, with **k** representing the wavevector, $\omega$ the angular frequency and $c$ the speed of light in vacuum. The $n_g$ value is calculated from the reciprocal of the slope of the guided even mode curve, and $n_g$ in the probe region is larger than in the pump region due to flatter dispersion. Since a high $n_g$ may lead to a large propagation loss, the pump wavelength should be selected in a region with a moderate $n_g$ to balance the loss and slow-light enhancement. In contrast, phase modulation is independent of the probe power, therefore the probe wavelength can be chosen in a region where a larger $n_g$ is preferable.

Fig. 2(b) depicts the calculated heat source $\tilde{Q}$, phase modulation $\Delta\tilde{\phi}$, $n_{gp}$ and $n_{gb}$ values as functions of wavelength. The $n_{gp}$ varies from 5 to 24 in the pump region while the $n_{gb}$ from 24 to 270 in the probe region. The change in $\tilde{Q}$ results from the combined effect of gas absorption coefficient and $n_g$-enhanced heating power. $\Delta\tilde{\phi}$ increases with rising wavelength, which is highly consistent with the trends of $n_g$ change against the wavelength, suggesting that higher $n_g$ enhances the PT-induced phase modulation. Fig. 2(c) shows the calculated change of temperature for three different absorption (pump) wavelength $\lambda_p$, corresponding to $n_{gp}$ of 5, 10 and 24, respectively. The peak temperature variation ($\tilde{T}$) increases linearly with rising $n_{gp}$ values. Fig. 2(d) shows the frequency-domain $\Delta\tilde{\phi}$ for three different probe wavelength $\lambda_b$, corresponding to $n_{gb}$ of 27, 58 and 106, respectively. $\Delta\tilde{\phi}$ also increases linearly with rising $n_{gb}$ values. Owing to the suspended design, the modulation bandwidth extends to ~1 MHz, which is beneficial for phase-sensitive detection at higher frequencies with lower 1/$f$ noise.[27]

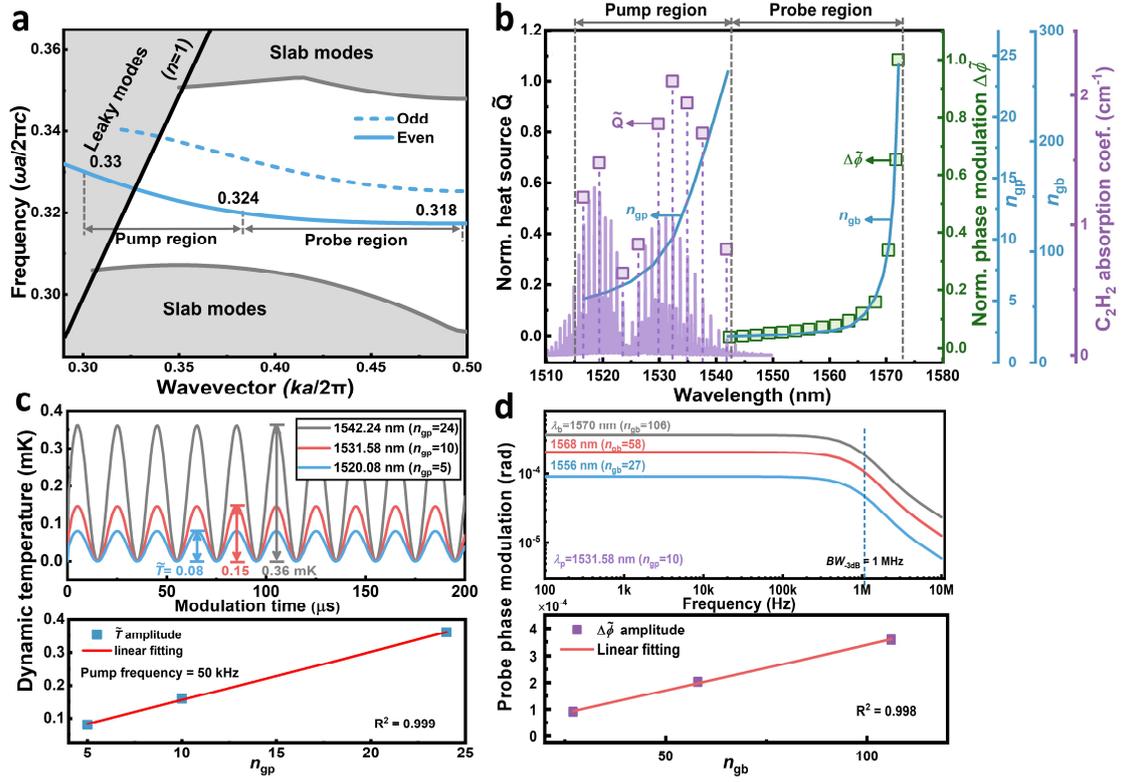

**Fig. 2 Numerical simulation of PTS based on PhCW. (a)** Dispersion diagram of the 2D-PhCW with optimized parameters. The slow-light wavelength band includes pump and probe regions. **(b)** The heat source $\tilde{Q}$, phase modulation amplitude $\Delta\tilde{\phi}$, $n_{gp}$ and $n_{gb}$ as functions of the wavelength. In the pump region, the values of $\tilde{Q}$ at the $C_2H_2$ absorption line centers are shown as the purple squares while $n_{gp}$ as a function of $\lambda_p$ is shown as the left blue line. In the probe region, $\Delta\tilde{\phi}$ (green squares, normalized values against their maxima) and $n_{gb}$ (right blue line) as functions of $\lambda_b$ are shown. **(c)** Time-domain temperature response for three different $\lambda_p$ of 1520.08, 1531.58, and 1542.24 nm, corresponding to $n_{gp}$ of 5, 10 and 24, respectively, at the pump modulation frequency of 50 kHz (upper panel). The peak temperature variation shows a linear relationship to $n_{gp}$ (lower panel). **(d)** Frequency-domain phase modulation $\Delta\tilde{\phi}$ for three different $\lambda_b$ of 1556, 1568 and 1570 nm, corresponding to $n_{gb}$ of 27, 58 and 106, respectively (upper panel). The $\Delta\tilde{\phi}$ increases linearly with rising $n_{gb}$ for a fixed $n_{gp}$ of 10 (lower panel).

**The PTS chip.** The silicon-integrated PTS chip is fabricated with a CMOS-compatible process (Details in Methods and Supplementary S3). As shown in Fig. 3(a), the device consists of an input/output subwavelength grating coupler (SWGC) and a Mach-Zender interferometer (MZI) with an ultracompact footprint of ~0.6 mm$^2$ (4.5L×0.13W mm$^2$). The strip waveguide after the input SWGC is narrowed down to a width of 0.5 μm over a length of 100 μm to ensure fundamental TE transmission in the MZI (Fig. S4). The waveguide width is then expanded to ~2 μm before the Y-splitter of the MZI. The suspended 2D-PhCW with a length of 1 mm forms part of the sensing arm. The reference arm consists of a single-mode strip waveguide with a width of 0.5 μm. As shown in Fig. 3(b), the Y-splitter features a waveguide width ratio of 3:1 (*i.e.,* power splitting ratio of 9:1) to compensate for the relatively larger loss of the PhCW and ensure approximately equal optical power in the two arms.[28] A Y-combiner is used to combine the fields from the two arms into the output SWGC for interferometric measurement. The corresponding SEM images of the 2D-PhCW and the suspended structure are shown in Figs. 3(c) and (d), respectively. A 100-μm-long tapered strip waveguide is fabricated before the PhCW to ensure precise excitation of the $TE_0$ mode, thereby selectively exciting TE even mode in the 2D-PhCW.[29-31] Fig. 3(e) shows more details of the transition region showing the gradual variation of hole size along the central line defect, which are used to reduce the coupling loss between the strip waveguide and the PhCW.

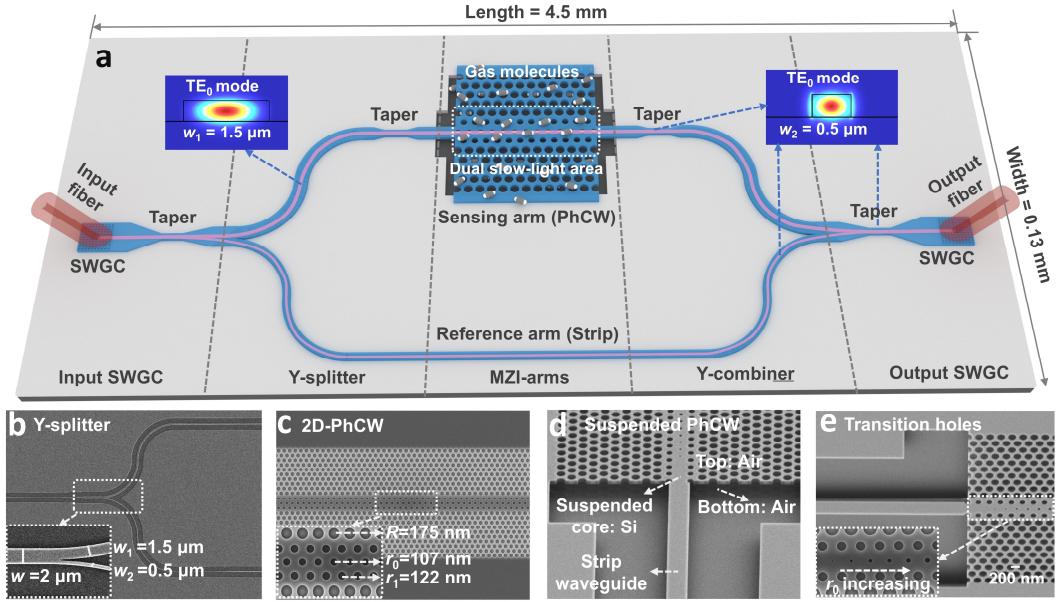

**Fig. 3 The PTS chip. (a)** The silicon-integrated MZI chip with a total length of 4.5 mm and a width of 0.13 mm. A 1-mm-long 2D-PhCW is located in the sensing arm and a single-mode strip waveguide serves as the reference arm. The two insets show the excited $TE_0$ modes with different widths of $w_1$=1.5 μm (sensing arm) and $w_2$=0.5 μm (reference arm). SEM images of the fabricated silicon chip include: **(b)** Y-power splitter with unequal width (1.5 and 0.5 μm) to balance the power distribution of the two MZI arms. **(c)** 2D-PhCW features three kinds of defect holes ($R$, $r_0$, $r_1$), with radii of $R$ = 175 nm, $r_0$ = 107 nm and $r_1$ = 122 nm. **(d)** Suspended structure of the 2D-PhCW with air serving as the top and bottom claddings, while silicon serving as the core layer. The strip waveguide near the PhCW has a suspended section of ~20 μm, while most of the strip waveguide are non-suspended. **(e)** SEM image of the transition region shows the defect holes (6 of 14). The radius of the defect holes increases gradually from 24 to 107 nm throughout the transition region (Fig. S5).

**Experimental characterization.** The PhCW for slow-light enhancement is characterized by its $n_g$ values, which can be determined from the wavelength difference ($\Delta\lambda$) between adjacent peaks of the on-chip MZI transmission (Details in Methods).[32] As shown in Fig. 4(a), in the spectral range of interest (1480−1575 nm), $\Delta\lambda$ reduces significantly from ~4.6 nm at 1480 nm to around ~0.02 nm at 1573 nm, and becomes small beyond 1573 nm, making it hard to determine the $n_g$ value accurately due to the background noise and the resolution of the optical spectrum analyzer (0.02 nm). Fig. 4(b) highlights the $n_g$ values deduced from the averaged transmission spectrum of adjacent two interference fringes. For wavelengths below 1572 nm ($n_g \approx 81$), the measurement is considered as reasonably accurate as $\Delta\lambda$ is larger than the spectral resolution. For wavelengths at 1573.5 nm ($n_g \approx 107$) and beyond, the measurement error may increase considerably due to the limited spectral resolution. Therefore, we choose probe wavelength at 1572 nm in our gas detection experiments, instead of a longer wavelength (corresponding to larger $n_g$) to enhance more of the PTS signal.

We then examine the slow-light enhanced pump absorption with a fixed probe wavelength (Detailed set-up in next section). Fig. 4(c) shows the measured PTS signal ($S$) of 5% $C_2H_2$ for 14 different wavelengths from ~1512 to ~1542 nm, corresponding to peaks of the $C_2H_2$ absorption lines. Here $S$ represents the first-harmonic ($1f$) amplitude demodulated from the MZI output. We compare the PTS performance with pump wavelengths set at point A and B. Although the absorption coefficient at point B (1.05 cm$^{-1}$ at 1531.58 nm) is ~90% that at point A (1.17 cm$^{-1}$ at 1520.08 nm), the PTS signal at point B is 1.74 times larger due to a higher value of $n_{gp}$ (1.95), which is highly consistent with the product (1.75) of the absorption coefficient and $n_{gp}$, indicating the slow-light enhanced pump absorption.

The slow-light enhancement on probe phase modulation is further investigated with pump wavelength fixed at B with 1% $C_2H_2$ filled into the gas cell. Fig. 4(d) shows the measured $S$, fringe contrast ($v$), and resultant $\Delta\tilde{\phi}$ at different probe wavelengths. The solid lines are the corresponding calculated values (Details in Supplementary S4). As probe wavelength increases, $\Delta\tilde{\phi}$ also rises due to the increment of $n_{gb}$, indicating that the slow-light effect can indeed enhance the probe phase modulation. However, with increasing $n_{gb}$, the slow-light-induced propagation loss in PhCW-based sensing arm increases correspondingly (Supplementary S9), which leads to a decreasing $v$ due to the unbalanced power of the two MZI arms.[32] Using a variable on-chip attenuator in the reference arm could rebalance the power distribution to further enlarge the MZI contrast and the PTS signal.[33] Overall, the dual slow-light enhancement is determined by both $n_{gp}$ and $n_{gb}$ values. Fig. 4(e) delves into the linear relationship between $\Delta\tilde{\phi}$ and the product of $n_{gp}$ and $n_{gb}$ ($R^2$=0.96), unveiling the dual slow-light enhancement in the pump-probe PTS scheme (Details in Supplementary S5).

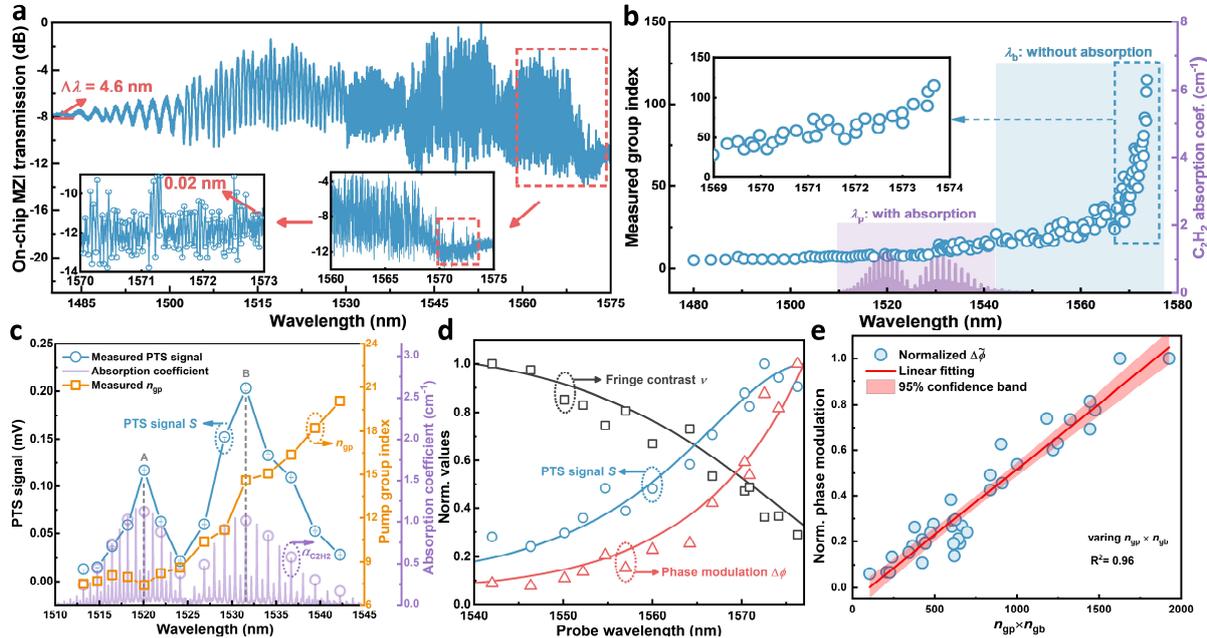

**Fig. 4 Experimental characterization of the dual slow-light PhCW PTS. (a)** Transmission spectrum of the on-chip MZI. The $n_g$ value is determined from the $\Delta\lambda$ between adjacent peaks. The insets show the enlargements within the wavelength ranges of 1560−1575 nm and 1570−1573 nm, respectively. **(b)** The $n_g$ values deduced from the averaged transmission spectrum of adjacent two interference fringes. The purple-shaded area with modest $n_{gp}$ values is the pump wavelength band covering the $C_2H_2$ absorption lines. The blue-shaded area has large $n_{gb}$ values, which is away from $C_2H_2$ absorption and serves as the probe wavelength band. **(c)** Measured PTS signal $S$ (blue circles) and $C_2H_2$ absorption coefficient (purple lines) as functions

of pump wavelength. The probe wavelength is fixed at 1572 nm. The local absorption coefficient maxima at A (1520.08 nm) and B (1531.58 nm) are selected for comparison. The orange square denotes $n_{gp}$ values. Error bars of the PTS signal show the standard deviation from five measurements. **(d)** Measured PTS signal $S$ (blue circles), fringe contrast $v$ (black squares), and deduced phase modulation amplitude $\Delta\tilde{\phi}$ (red triangles) as functions of probe wavelength with the pump wavelength fixed at 1531.58 nm. **(e)** The measured $\Delta\tilde{\phi}$ as a function of $n_{gp} \times n_{gb}$ normalized against its maxima. The red line is the linear fitting with $R^2$ value of 0.96.

**On-chip gas sensing.** The performance of dual slow-light enhanced PTS for gas sensing is evaluated by utilizing the set-up shown in Fig. 5(a). The pump is from a distributed feedback (DFB) laser, with its wavelength fixed at the center of the P(11) absorption line of $C_2H_2$ at 1531.58 nm. The pump intensity is modulated by an acousto-optic modulator (AOM) and amplified by an Er-doped fiber amplifier (EDFA). The probe beam, from an external cavity diode laser (ECDL) with a 100 kHz linewidth, is co-coupled with the pump beam into the on-chip MZI via a wavelength-division multiplexer (WDM). Two polarization controllers (PC) are used to adjust the polarization state of the pump and probe to maximize the fiber-SWGC coupling efficiency. The MZI output is filtered by a narrowband filter (<1 nm) to block the residual pump beam, and the transmitted probe beam is focused onto a photodiode (PD). The PD signal is sent to a lock-in amplifier (LIA) for 1$f$ demodulation. A sinusoidal signal from a function generator (FG) drives the AOM, which also serves as a reference for the LIA. The silicon-integrated on-chip MZI with the sensing PhCW is packaged inside a compact gas cell with a dimension of 5×2×1 cm$^3$ and a volume of <10 mL (Supplementary S6). Gas sensing experiments are conducted by filling different $C_2H_2$:$N_2$ mixtures into the packaged gas cell.

Fig. 5(b) shows the 1$f$-PTS signal and the 1$\sigma$ noise level as functions of the pump power coupled into the PhCW. We examined the PTS signal and noise with two pairs of pump-probe wavelengths. The first pair has a pump wavelength of 1531.58 nm and a probe wavelength of 1572 nm, corresponding to a large $n_{gp} \times n_{gb}$ value of 1458. In contrast, the second pair utilizes a pump wavelength of 1520.08 nm and a probe wavelength of 1542 nm with a relatively small $n_{gp} \times n_{gb}$ value of 176. The PTS signals for both pairs show approximately linear relationships with the pump power, while the amplitude in the first pair is ~8 times larger than the latter, which is close to the proportion of their $n_{gp} \times n_{gb}$ values. There is no significant increase in noise levels with increasing $n_{gp} \times n_{gb}$ values, indicating the signal-to-noise ratio (SNR) can be enhanced by operating at the slow-light regime (Supplementary S7). Fig. 5(c) shows the measured PT efficiency with varying pump modulation frequency, which exhibits no significant change at low frequencies (<100 kHz) and decreases relatively rapidly beyond 500 kHz. The $k^*$ value at 50 kHz is 3.6×10$^{-4}$ rad·cm·ppm$^{-1}$·mW$^{-1}$·m$^{-1}$ and the 3−dB modulation bandwidth is ~500 kHz (Details in Methods). Fig. 5(d) shows the linear relationship of the 1$f$-PTS signal with increasing $C_2H_2$ concentration from 0 to 30%, giving a linear dynamic range of >2.5×10$^4$. While the PTS signal scales linearly with concentration, the noise level remains stable, leading to a proportional increase in SNR across entire dynamic range. As a result, the precision remains approximately unchanged across different concentration levels.

Fig. 5(e) shows the Allan-Werle analysis based on the measured noise level over a 2-h period with pump wavelength tuned away from the absorption line at 1531.27 nm. As shown in Figs. S15(a) and (b), two more long-term noise measurements with pure $N_2$ were conducted under different conditions: when the pump was aligned to the absorption line at 1531.58 nm and away from absorption at 1531.27 nm, respectively. The $n_{gp}$ value at these two wavelengths remains essentially the same. The Allan deviation results indicate that at an averaging time of ~200 s, the 1σ minimum detection limit (MDL) reaches its minimum in both cases: 15 ppm (at 1531.58 nm) and 12 ppm (at 1531.27 nm ), corresponding to a minimal noise-equivalent absorption (NEA) coefficient of $1.4\times10^{-5}$ $cm^{-1}$. We further evaluated the signal fluctuation at the $C_2H_2$ concentration of 1000 ppm, as shown in Fig. S15(c). The 1σ fluctuation is estimated to be 18 ppm at an averaging time of 100 s, slightly larger than the 1σ MDL value obtained with pure $N_2$ (Supplementary S10). With respect to the sources of noise and drift over extended measurement times, at short integration time within 200 s, the noise mainly emanates from the relative intensity noise of the probe laser and electronic noise of the electrical circuits. At longer time over 200 s, the system drift dominates, which is mainly due to external temperature variation and pump/probe light power fluctuations that vary slowly over time.

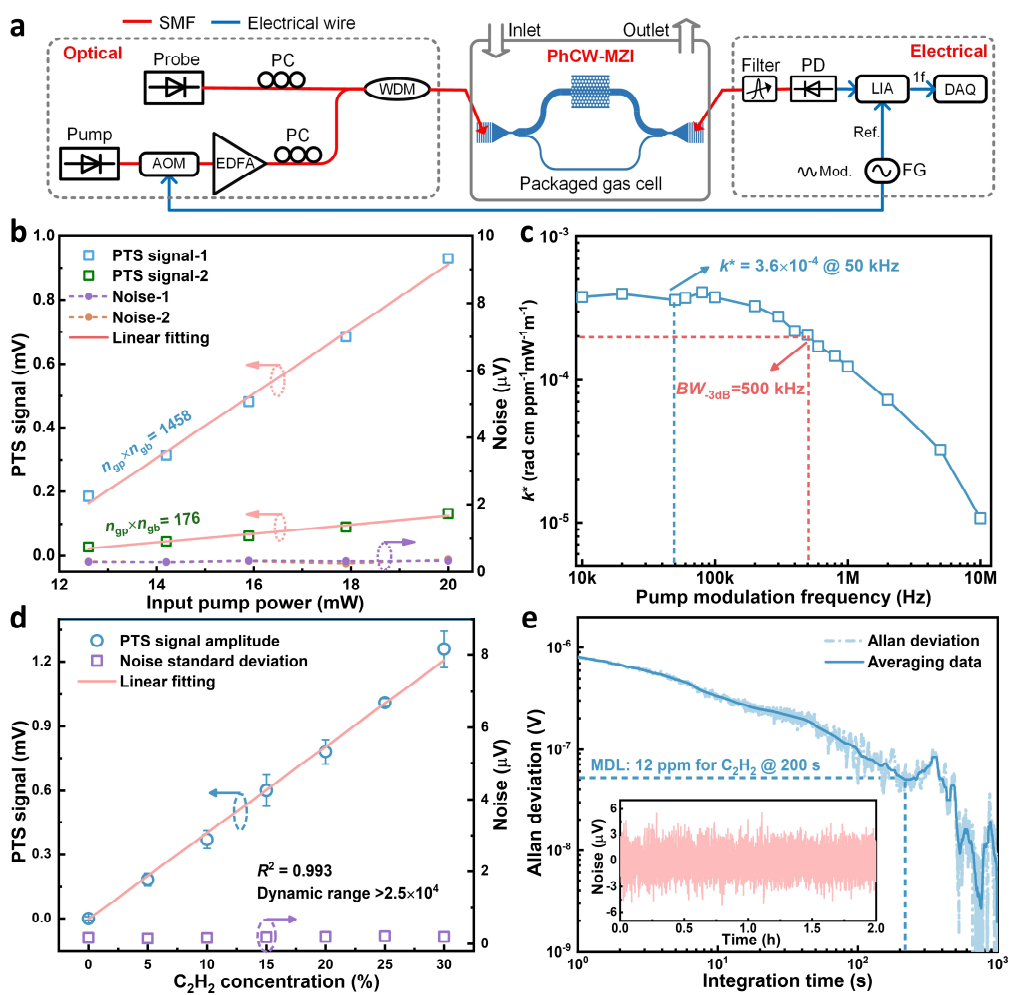

**Fig. 5 Dual slow-light enhanced PTS for on-chip gas sensing. (a)** Experimental set-up. AOM: acousto-optic modulator, EDFA: Er-doped fiber amplifier, PC: polarization controller, WDM: wavelength division multiplexer, FG: function generator, PD: photodetector, LIA: lock-in amplifier, DAQ: data acquisition card. The operation point of the on-chip MZI is sufficiently stable to obviate the need of external electronic feedback locking (Supplementary S6). **(b)** $1f$-PTS signal and the standard deviation of the noise as functions of the input pump power for pump modulation $f$ = 50 kHz. The $C_2H_2$ concentration is $C$ = 5%. The results are obtained for two pairs of pump and probe wavelengths, corresponding to $n_{gp}\times n_{gb}$ = 1458 and 176, respectively. **(c)** Measured $k^*$ value against pump modulation frequency. **(d)** PTS signal and noise level as functions of $C_2H_2$ concentration from 0 to 30% ($P_p$ = 12.5 mW, $f$ = 50 kHz). Error bars of the

PTS signal show the standard deviation from five measurements. **(e)** Allan deviation plot based on the recorded noise data (inset), along with the averaged data using smoothing method. The LIA time constant is set as 100 ms, corresponding to 0.94 Hz detection bandwidth.

## Discussion

In summary, we have demonstrated a scheme of dual slow-light enhancement in PTS sensing on a suspended silicon PhCW. Based on Maxwell's eigenvalue equation, our perturbation theory demonstrated that the dual slow-light effect originates from the reduced energy transport velocity in the PhCW, where the subsequent perturbation on whether the real or imaginary part of the eigen value can be amplified with respect to the $n_g$ values. By using the approximate solution to the heat transfer, we proved that both pump absorption and probe phase modulation can be enhanced within tailored PhCWs by the structural dual slow-light effect, resulting in a dramatic increase of the PT efficiency. The experimental results confirmed the enhancement of PT phase modulation by $n_{gp} \times n_{gb}$ times, consistent with our theoretical formulations. Since the slow-light effect is directly related to the slope of the dispersion curve of the PhCW mode, it would be possible to further optimize the PhCW to achieve higher $n_{gp} \times n_{gb}$ values.

The PT efficiency can be quantified by its $k^*$ value, which is used to compare the performance of PTS with different optical fibers and waveguides. Previous work reveals that the silica HCF has a $k^*$ value of $\sim 8 \times 10^{-7}$ rad·cm·ppm$^{-1}$·mW$^{-1}$·m$^{-1}$ at 19 kHz.[17] The chalcogenide strip waveguide shows a relatively high $k^*$ value of $3.05 \times 10^{-5}$ rad·cm·ppm$^{-1}$·mW$^{-1}$·m$^{-1}$ at 6 kHz owing to the higher TO coefficients of waveguide materials as compared to the gas medium.[23] The dual slow-light PhCW in this work achieves a $k^*$ value of $\sim 3.6 \times 10^{-4}$ rad·cm·ppm$^{-1}$·mW$^{-1}$·m$^{-1}$ at 50 kHz, outperforming the conventional HCFs by nearly 3 orders of magnitude and the waveguides by over an order of magnitude. Such a high $k^*$ coefficient is also attributed to the suspended structure that enhances heat accumulation (Supplementary S8).

We also conducted on-chip direct absorption spectroscopy (DAS) using a 1-mm-long suspended 2D-PhCW, which involves a single slow-light enhancement (Supplementary S9). By keeping the same $n_{gp}$ value for both DAS and PTS, DAS demonstrates a $C_2H_2$ detection limit of 436 ppm at 45 s, nearly 2 orders of magnitude inferior to that of dual slow-light enhanced PTS. This improvement in PTS is primarily due to the enhanced phase modulation from the $n_{gb}$ value and the low-noise performance of PTS over DAS.[23] We further compared the performance of our dual slow-light PTS sensor with other photonic waveguide sensors in terms of NEA·L[34], as listed in Table 1. Detailed parameters are provided in Tables S1 and S3 (Supplementary S11). The results reveal that the NEA·L for our sensor is 1−3 orders of magnitude superior to the other reported photonic waveguide and DAS-based PhCW/SWG gas sensors. Furthermore, when compared to state-of-the-art free-space and fiber-based sensors, as listed in Table S2, the NEA·L sensitivity of our sensor approach that of the advanced free-space/fiber sensors, while offering distinct benefits in compactness, CMOS compatibility, and scalable integration, which are key features for miniaturized and distributed gas sensing platforms.

We note that the MDL for $C_2H_2$ detection remains relatively higher than that of mid-infrared photonic sensors.[35,36] However, the achieved NEA·L sensitivity is impressive, and the fact that this is achieved in the near-infrared band, where photonics is mature and easy integrated with both light sources and detectors, is of high value. The main limitations of the current PTS device stem from its optical losses. The slow-light-induced propagation loss increases as $n_g$ increases, which is an intrinsic property in PhCW-PTS devices. For the pump beam, this can be overcome by increasing the laser power. For the probe beam, if the light level is sufficient at the PD, it would not be a problem. Another waveguide-based propagation loss is mainly due to the sidewall roughness and could be minimized by optimizing the lithography and etching fabrication process. The fiber-SWGC coupling loss of ~9 dB/facet is relatively large for this device, which could be reduced to below ~1 dB by optimizing the SWGC design.[37] In addition, using a SOI wafer with a thicker 3-μm buried oxide layer would help to reduce possible substrate leakage at the working wavelengths.

The dual slow-light enhanced PTS sensor shows high performance in sensitivity and footprint, which would pave the way for full integration into miniaturized silicon photonic circuits. We believe the innovative dual slow-light scheme is also applicable to high-performance liquid sensors,[38] mid-infrared photonic sensors by taking advantage of the strong fundamental molecular absorption,[39] and other dual-light systems for the study of light-matter interactions, with potential applications in smart homes,[40] point-of-care diagnostics,[41] and wearable devices.[42,43]

**Table 1** Performance comparison of photonic waveguide gas sensors

| Methods | Gases & $\lambda$ (nm) | Wavelengths | Chip types and lengths | NEA·L |
|---|---|---|---|---|
| DAS[44] | $CH_4$ @1650 nm | NIR | 10 cm Si-strip | $4.6\times10^{-4}$ |
| DAS[45] | $C_2H_2$ @2566 nm | MIR | 2 cm $Ta_2O_5$-strip# | $3.6\times10^{-4}$ |
| DAS[36] | $CH_4$ @3270 nm | MIR | 1.15 cm Si-Slot | $1.7\times10^{-5}$ |
| DAS[46] | $CO_2$ @4240 nm | MIR | 0.32 cm Si-strip# | $9.9\times10^{-2}$ |
| DAS[35] | $CO_2$ @4345 nm | MIR | 2 cm SiN-PhCW# | $6.8\times10^{-6}$ |
| DAS[47] | $C_7H_8$ @6650 nm | MIR | 2.84 cm Si-SWG | $9.6\times10^{-4}$ |
| DAS[48] | $C_3H_6O$ @7330 nm | MIR | 1 cm Si-SWG | $1.1\times10^{-4}$ |
| Off-chip PTS[23] | $C_2H_2$ @1531 nm | NIR | 2 cm ChG-strip | $9.3\times10^{-6}$ |
| Off-chip PTS[49] | $CO_2$ @2004 nm | NIR | 9.12 cm LN-strip | $1.1\times10^{-3}$ |
| On-chip PTS[This] | $C_2H_2$ @1531 nm | NIR | 0.1 cm Si-2D-PhCW# | $1.4\times10^{-6}$ |

NEA·L is independent of the absorption line strength and interaction length (L). NIR: Near-infrared, MIR: Mid-infrared. Si: Silicon, $Ta_2O_5$: Tantalum pentoxide, SiN: Silicon nitride, ChG: Chalcogenide glass, LN: Lithium niobate, SWG: Sub-wavelength grating, $CH_4$: Methane, $CO_2$: Carbon dioxide, $C_7H_8$: Toluene, $C_3H_6O$: Acetone. #Suspended waveguide structure with air as top and bottom claddings. Off-chip (fiber-based) interferometers are used for PTS with complicated servo-control to lock the interferometer at quadrature.[23,49] This on-chip MZI has an ultrasmall footprint of only 0.6 mm$^2$ and no servo-locking device is needed (Supplementary S6).

# Methods

**Numerical simulation.** The photonic band structure and dispersion relation of the PhCW are tailored and simulated using the RSoft band-solver. The group index is extracted from the dispersion diagram. Each band is fitted based on 100 points in the lattice along the Γ-K direction. The slow-light parameter optimizations and transmission modes of the PhCW are simulated using Ansys Lumerical FDTD. The optical mode and thermal field as well as time- and frequency-domain profile are simulated using a three-dimensional finite-element-method (3D-FEM) with COMSOL Multiphysics. For the optical mode calculation, the pump and probe modes are calculated using eigen solver in COMSOL with the Floquet boundary condition applied to the front and back surface. For thermal field, a symmetry boundary is applied to the front and back surfaces to simulate a scenario where the structure is infinitely periodic along the $z$-axis. The other boundaries surrounding the PhCW are allowed to transfer the heat to the air via natural convective heat flux. The phase modulation $\Delta\tilde{\phi}$ is calculated by integrating the mode overlapping between the probe and the thermal fields over the entire PhCW structure.

**Silicon chip fabrication.** The silicon-based photonic crystal patterns of the 2D-PhCW, input/output SWGCs, and other strip waveguides in the MZI are fabricated on the same SOI wafer using electron beam lithography (EBL) and inductively coupled plasma (ICP) dry-etching. The silica bottom cladding of the photonic crystal region is wet-etched subsequently using buffer oxide etching solution. Before wet-etching, ultraviolet (UV) photoresist is used to protect the strip waveguide and SWGC areas, preventing collapse due to excessive etching. A suspended 2D-PhCW is obtained after removing the UV photoresist.

**Group index determination.** To determine the $n_g$ values of the slow-light PhCW, the on-chip integrated MZI is connected with a broadband light source and an optical spectrum analyzer (OSA). The broadband source encompasses a wavelength range of 1450–1640 nm, facilitated by an OSA with a spectral resolution of 20 pm. The $n_g$ values of the PhCW can be derived from the interference spectrum using the following formula $n_g(\lambda) = \lambda_1 \times \lambda_2/(\Delta\lambda \times L) + n_g^{ref}(\lambda)$, where $\lambda_1$ and $\lambda_2$ represent the wavelengths of the fringe peaks, $\Delta\lambda = |\lambda_1 - \lambda_2|$ the wavelength difference between adjacent peaks, $L$ the PhCW length, and $n_g^{ref}(\lambda)$ is the group index of the reference arm.[5]

**PT efficiency determination.** The PT efficiency $k^*$ value of the dual slow-light enhanced PTS on PhCW is determined based on the experimental result using formula $k^* = \Delta\tilde{\phi}/\alpha C P_p L$.[17] Here $\Delta\tilde{\phi}$ is obtained by comparing the experimental 1$f$-LIA output with fringe contrast. $\alpha$ is the absorption coefficient of the P(11) rovibrational transition, which is taken from the HITRAN database. $C$ is the gas concentration, $P_p$ the input pump power, and $L$ is the PhCW length. Under 5% $C_2H_2$ concentration and 1-mm-length PhCW, a $k^*$ value of $3.6 \times 10^{-4}$ rad·cm·ppm$^{-1}$·mW$^{-1}$·m$^{-1}$ is obtained at the modulation frequency of 50 kHz.

## Acknowledgements

The authors wish to express their gratitude to the National Natural Science Foundation of China (62175087, C.Z.; 62235016, C.Z.; 62535009, C.Z.), the Hong Kong SAR government GRF (15223421, W.J.), the Key Science and Technology R&D program of Jilin Province, China (20230201054GX, C.Z.), the Local Innovative and Research Teams Project of Guangdong Pearl River Talents Program (2019BT02X105, W.J.), and the Hong Kong PolyU (W23D, W.J.; W229, W.J.; CDJ6, W.J.). We also wish to acknowledge Lifu Duan for the helpful discussions.

## Author contributions

C.Z. conceived the initial idea. K.Z. designed and performed the experiments. Z.P. designed and optimized the PhCW. H.L. developed the mathematical formulations. K.Z., Z.P. and H.L. performed the numerical calculation. K.Z. analyzed the results and prepared the manuscript. Z.P. and H.L. assisted in preparing the manuscript. Y.H. assisted in performing the numerical calculation. H.B., S.Z., Y.Z. and Y.W. assisted in analyzing the results. C.Z. and W.J. coordinate the project. All authors reviewed the manuscript and provided editorial input.

## Data availability

All information necessary to support the findings of this study is available in the main article and Supplementary Information file.

## Competing interests

The authors declare no competing interests.

## Additional information

**Supplementary information**. The online version contains supplementary material.

**Correspondence and requests for materials** should be addressed to Chuantao Zheng or Wei Jin.

**Reprints and permissions information** is available at www.nature.com/reprints.

# Supplementary Information for

# Dual slow-light enhanced photothermal gas spectroscopy on a silicon chip

Kaiyuan Zheng,[1,2,3,†,#] Zihang Peng,[1,2,3,†] Hanyu Liao,[2,3,†] Yijun Huang,[1] Haihong Bao,[2,3] Shuangxiang Zhao,[2,3] Yu Zhang,[1] Chuantao Zheng,[1,*] Yiding Wang,[1] Wei Jin[2,3,*]

[1]State Key Laboratory of Integrated Optoelectronics, College of Electronic Science and Engineering, Jilin University, 2699 Qianjin Street, Changchun, 130012, China
[2]Department of Electrical and Electronic Engineering and Photonics Research Institute, The Hong Kong Polytechnic University, Hung Hom, Hong Kong, 999077, China
[3]Photonics Research Center, The Hong Kong Polytechnic University Shenzhen Research Institute, 18 Yuexing Road, Shenzhen, 518071, China
[†]These authors contributed equally to this work
[#]Present address: Division of Environment and Sustainability, The Hong Kong University of Science and Technology, Clear Water Bay, Hong Kong, 999077, China
[*]Corresponding authors: zhengchuantao@jlu.edu.cn, wei.jin@polyu.edu.hk

## S1. Theory of the dual slow-light enhanced PTS

We describe the principle of dual slow-light enhancement in photonic crystal waveguide (PhCW)-based photothermal spectroscopy (PTS) using perturbation theory. Firstly, we show that any linear perturbation on the material permittivity results in a proportional change in the optical mode refractive index (RI) scaled by the group index. Then, we treat both pump absorption and probe phase modulation as perturbations on the permittivity, which are accordingly enhanced by the group index. The only difference is that pump absorption introduces a perturbation to the imaginary part of the RI, while probe phase modulation introduces a perturbation to the real part of the RI. Finally, we derive the PT phase modulation induced by pump absorption on the probe beam, revealing that the high group index of PhCW leads to a dual enhancement in the PT phase modulation.

### *1.1 Perturbation theory on mode equation*

We consider Maxwell's equations as an eigen problem, where the electric field is expressed in the Dirac notation $|E\rangle$ with time dependence $\exp(-i\omega t)$. For a source-free linear dielectric $\varepsilon$, the eigen mode equation for the electric field is written as:[1]

$$\nabla \times \nabla \times |E\rangle = \left(\frac{\omega}{c}\right)^2 \varepsilon |E\rangle \tag{S1}$$

where $\omega$ is the optical angular frequency, $c$ the speed of light, $\varepsilon = \epsilon + i\epsilon`$ the permittivity, and the inner product of the eigen modes is $\langle E|E`\rangle = \int \boldsymbol{E}^* \cdot \boldsymbol{E}\, \mathrm{d}V$. Using the first-order perturbation theory, $\varepsilon = \varepsilon^{(0)} + \delta\varepsilon$, the perturbation on the eigen solution gives $|E\rangle = |E^{(0)}\rangle + |E^{(1)}\rangle$ with its eigen value $\omega = \omega^{(0)} + \delta\omega$. The zeroth order equation and the first-order equation are accordingly written as:

$$\begin{cases} \nabla \times \nabla \times |E^{(0)}\rangle = \left(\frac{\omega^{(0)}}{c}\right)^2 \varepsilon^{(0)} |E^{(0)}\rangle \\ \nabla \times \nabla \times |E^{(1)}\rangle = \frac{(\omega^{(0)})^2}{c^2} \varepsilon^{(0)} |E^{(1)}\rangle + \frac{(\omega^{(0)})^2}{c^2} \delta\varepsilon |E^{(0)}\rangle + \frac{2\omega^{(0)}\delta\omega}{c^2} \varepsilon^{(0)} |E^{(0)}\rangle \end{cases} \tag{S2}$$

Using the orthogonality condition $\langle E^{(0)}|\varepsilon^{(0)}|E^{(1)}\rangle = 0$, the first-order equation could be simplified to:

$$\delta\omega = -\frac{\omega^{(0)}}{2}\frac{\langle E^{(0)}|\delta\varepsilon|E^{(0)}\rangle}{\langle E^{(0)}|\varepsilon^{(0)}|E^{(0)}\rangle} \tag{S3}$$

Eq. (S3) shows that the perturbation on permittivity changes the eigenfrequency of the optical mode, which should correspond to a correction of the propagation wave vector $\delta\mathbf{k}$. The relationship between the optical frequency $\omega$ and the propagation constant $k$ is given by the dispersion relation of the PhCW in the form of $D(\omega, \mathbf{k}, \mu) = 0$, where $\mu$ denotes the material constant. Under an arbitrary but small perturbation of the material permittivity $\delta\varepsilon$, the perturbed dispersion relation can be written as:[2]

$$D(\omega, \mathbf{k}, \varepsilon + \delta\varepsilon) = D(\omega - \delta\omega|_{\mathbf{k}^{(0)}}, \mathbf{k}, \varepsilon) = D(\omega, \mathbf{k} - \delta\mathbf{k}|_{\omega^{(0)}}, \varepsilon) \tag{S4}$$

Under the first-order perturbation assumption, around any point $(\omega^{(0)}, \mathbf{k}^{(0)})$ of the dispersion relation where the group velocity does not vanish, the dispersion relation can be written in a localized form:[2]

$$D(\omega, \mathbf{k}, \varepsilon) = \omega - \omega^{(0)} - v_g(\mathbf{k} - \mathbf{k}^{(0)})^1 = 0 \tag{S5}$$

$$v_g = -\frac{\partial D}{\partial \mathbf{k}}\bigg/\frac{\partial D}{\partial \omega} \tag{S6}$$

where $v_g$ is the group velocity in the absence of loss. Therefore, Eq. (S4) becomes:

$$\delta\omega = -v_g\delta\mathbf{k} = -\frac{c}{n_g}\cdot\left(\frac{\omega}{c}\delta n_m\right)\bigg|_{\omega=\omega^{(0)}} \tag{S7}$$

here $n_g = c/v_g$ is the group index. Combining the above dispersion relation with Eq. (S5), we can get:

$$\delta n_m = -\frac{n_g}{\omega^{(0)}}\delta\omega = \frac{n_g}{2}\frac{\langle E^{(0)}|\delta\varepsilon|E^{(0)}\rangle}{\langle E^{(0)}|\varepsilon^{(0)}|E^{(0)}\rangle} \tag{S8}$$

Eq. (S8) shows that the perturbation on the permittivity results in a change of mode RI scaled by the group index, which is the origin of the slow-light enhancement on light-matter interaction. It has been reported that only the structural slow light has an enhancement on the light-matter interaction, while the material slow light does not.[3] This difference is because the slow light enhancement in principle does not depend on the group velocity defined as $d\omega/d\mathbf{k}$, but the energy transport velocity of the electromagnetic field ($v_{EM}$), which is defined as:[4]

$$v_{EM} := \frac{\langle S \rangle_t}{\langle U \rangle_t} \tag{S9}$$

where $S$ is the Poynting vector, $U$ is the energy density, and the bra-ket with subscript $t$ denotes the time-averaged value. In the lossless periodic media, the structural slow light satisfies $v_{EM} = v_g$ and one may find it unnecessary to distinguish between them. However, in the presence of loss or disorder, the group velocity $d\omega/d\mathbf{k}$ may no longer relate to the dispersion relation $D(\omega, \mathbf{k}, \varepsilon)$, rendering Eq. (S6) incorrect. Nonetheless, the energy transport velocity essentially follows the shape of the dispersion relation $D$ for TEM mode propagation, given as:[5]

$$v_{\text{EM}} = -\frac{\partial D}{\partial \mathbf{k}} \bigg/ \frac{\partial D}{\partial \omega} = c\frac{2n}{1 + n^2 + (\sigma C)^2} \tag{S10}$$

where $\sigma C = Im(N)$ is defined through the complex reflective index (RI) $N = n + i\sigma C$. Eq. (S10) proves that the energy transport velocity is always smaller than the speed of light. For the acetylene (C$_2$H$_2$) absorption with typically $\sigma \approx 1.28 \times 10^{-11}$ ppm$^{-1}$, it also reduces the energy transport velocity but is sufficiently small to be neglected.

### *1.2 Slow-light enhancement on pump absorption*

The pump beam experiences gas absorption which gives an imaginary-valued perturbation of $\delta\varepsilon$. Considering the case of trace gas detection with a weakly absorbing gas sample, the absorption coefficient is defined as $\alpha C$, where $C$ is the gas concentration and $\alpha = 1.05$ cm$^{-1}$ for C$_2$H$_2$ P(11) absorption line. The corresponding perturbation on the gas RI is described as:

$$n_{\text{gas}} = n_{\text{gas}}^{(0)} + \delta n = n_{\text{gas}}^{(0)} + \frac{ic\alpha C}{2\omega} \tag{S11}$$

here $n_{\text{gas}}^{(0)}$ is the RI of gas without light absorption (*e.g.,* ~1.0003 for air). Using the relation $\varepsilon = \varepsilon_0 \varepsilon_r = \varepsilon_0 n^2$, we have $\delta\varepsilon = 2\varepsilon_0 n_{\text{gas}}^{(0)} \delta n$, where the perturbation on permittivity is given as:

$$\delta\varepsilon = 2\varepsilon_0 n_{\text{gas}}^{(0)} \delta n = 2\varepsilon_0 n_{\text{gas}}^{(0)} \frac{ic\alpha C}{2\omega} \tag{S12}$$

Substitute the perturbation term into Eq. (S8) with Eq. (S12), we get:

$$\delta n_{\text{m}} = \frac{n_g}{2} \frac{\left\langle E^{(0)} \left| 2n^{(0)} \frac{ic\alpha C}{2\omega} \varepsilon_0 \right| E^{(0)} \right\rangle_g}{\left\langle E^{(0)} | \varepsilon^{(0)} | E^{(0)} \right\rangle} = \frac{n_g}{n_{\text{gas}}^{(0)}} \frac{ic\alpha C}{2\omega} \frac{\left\langle E^{(0)} | \varepsilon^{(0)} | E^{(0)} \right\rangle_g}{\left\langle E^{(0)} | \varepsilon^{(0)} | E^{(0)} \right\rangle} := \frac{n_g}{n_{\text{gas}}^{(0)}} \frac{ic\alpha C}{2\omega} \gamma \tag{S13}$$

here $\gamma$ is defined as the spatial confinement factor, which is the integration of the energy density in gas $\langle U \rangle_g$ divided by the total energy density $\langle U \rangle$.[6] Since the absorption is only induced by the gas, we have reduced the integration in the numerator to the gas region only. Assuming that the change of the mode field distribution due to gas absorption can be neglected, the intensity change of the pump mode along the PhCW can be written as:

$$I = \frac{1}{2}c\varepsilon_0 \langle E|n|E\rangle \approx \frac{1}{2}c\varepsilon_0 \langle E^{(0)}|n^{(0)}|E^{(0)}\rangle \exp\left(-2\frac{\omega}{c} Im(\delta n_{\text{m}})z\right) = I^{(0)} \exp\left(-\frac{n_g}{n_{\text{gas}}^{(0)}} \alpha C\gamma z\right) \tag{S14}$$

here $I^{(0)} = I_0 \exp(-\alpha_L z)$ is the optical intensity along the PhCW with $\alpha_L$ representing the insertion loss. Eq. (S14) proves that the Beer-Lambert absorption of the pump beam is linearly proportional to the group index $n_g$.

### *1.3 Slow-light enhancement on probe phase modulation*

The pump absorption induces a heat source that heats up the PhCW and changes the RI of the material. The probe beam experiences the RI change which gives a real-valued perturbation of $\delta\varepsilon$. For simplicity, we consider the heat source $Q$ induced by the pump absorption of gas molecules under a two-level model:

$$Q + \tau_{\text{re}} \frac{dQ}{dt} = \alpha C I_g^{(0)} \tag{S15}$$

here $\tau_{\text{re}}$ is the collision relaxation time of the gas molecule rovibrational transition, $I_g$ is the optical intensity with subscript g denotes the distribution in gas. Since the pump beam is intensity modulated, Eq. (S15) can be solved in the frequency domain as:

$$\tilde{Q} = \frac{m\alpha C}{1 + i\omega_m \tau_{re}} I_g^{(0)} \tag{S16}$$

where "~" denotes the frequency-domain amplitude, $m$ the modulation depth. Here the heat source is in the unit of W/m³ and only non-zero in the gas domain. The power of the heat source is proportional to the confinement factor $\Gamma$ of the pump mode as:

$$P_Q = \langle \tilde{Q} \rangle \propto \alpha C \left\langle E^{(0)} \middle| \frac{1}{2} c\varepsilon_0 n^{(0)} \middle| E^{(0)} \right\rangle_g = \alpha C \Gamma P^{(0)} \tag{S17}$$

here $\Gamma$ is the total confinement factor and can be expressed based on Eq. (S9) as:[6]

$$\Gamma = \frac{\langle I_g^{(0)} \rangle}{P^{(0)}} = \frac{\langle I^{(0)} \rangle_g}{\langle S^{(0)} \rangle} = \frac{n_g}{n_{gas}^{(0)}} \frac{\langle U \rangle_g}{\langle U \rangle} = \frac{n_g}{n_{gas}^{(0)}} \frac{\langle E^{(0)} | \varepsilon^{(0)} | E^{(0)} \rangle_g}{\langle E^{(0)} | \varepsilon^{(0)} | E^{(0)} \rangle} = \frac{n_g}{n_{gas}^{(0)}} \gamma \tag{S18}$$

Again, $\gamma$ is defined as the spatial confinement of the energy in gas $\langle U \rangle_g$ normalized by the total energy $\langle U \rangle$. The eigen solution $|E\rangle$ in $\gamma$ may also be replaced by the normalized mode field distribution as $|\psi\rangle$ in the unit of 1/m for simplicity. For trace gas detection, the heating process is weak, and the heat transfer in PhCW can be analyzed as:

$$i\omega \rho_g c_g \tilde{T} + \nabla \cdot (-\kappa_g \nabla \tilde{T}) + \rho_g c_g \vec{u} \cdot \nabla \tilde{T} = \tilde{Q} \tag{S19}$$

$$i\omega \rho_{Si} c_{Si} \tilde{T} + \nabla \cdot (-\kappa_{Si} \nabla \tilde{T}) = 0 \tag{S20}$$

where $\rho, c, \kappa, \vec{u}$ are the density, heat capacity, heat conductivity, and velocity field, the subscripts g and Si represent gas and PhCW, respectively. The solution of Eq. (S19–S20) is the temperature field oscillating with an amplitude $\tilde{T}$, which leads to the modulation of the accumulated phase of the probe beam primarily via the thermo-optic (TO) effect. The perturbation on RI can accordingly be written as:

$$\delta n = \frac{dn}{dT} \tilde{T} \tag{S21}$$

Substitute Eq. (S21) into Eq. (S12), we get the perturbation on permittivity as:

$$\delta \varepsilon = 2\varepsilon_0 n^{(0)} \frac{dn}{dT} \tilde{T} \tag{S22}$$

Combining Eq. (S22) with Eq. (S8), the phase modulation due to the oscillating temperature field is then:

$$\Delta \tilde{\phi} = \frac{2\pi L}{\lambda^{(0)}} \delta n_m = \frac{2\pi n_g L}{\lambda^{(0)}} \frac{\left\langle E^{(0)} \middle| \varepsilon_0 n^{(0)} \frac{dn}{dT} \tilde{T} \middle| E^{(0)} \right\rangle}{\langle E^{(0)} | \varepsilon^{(0)} | E^{(0)} \rangle} \tag{S23}$$

Here $\Delta \tilde{\phi}$ is the probe phase modulation amplitude. Since the probe phase perturbation is a real number, the denominator in Eq. (S23) is simply a normalization of the probe mode energy $\langle U \rangle$ on the PhCW. Hence, we replace the electric field $|E^{(0)}\rangle$ by its normalized mode field distribution $|\psi\rangle$ similar to Eq. (S18). Eq. (S23) can be accordingly simplified as

$$\Delta \tilde{\phi} = \frac{2\pi n_g L}{\lambda^{(0)}} \left\langle \psi \middle| \frac{dn}{dT} \tilde{T} \middle| \psi \right\rangle = \frac{2\pi n_g L}{\lambda^{(0)}} \left( \frac{e_{Si}^{TO}}{n_{Si}} \langle \psi | \tilde{T} | \psi \rangle_{Si} - \frac{\mu p}{\bar{T}^2} \frac{1}{n_{gas}^{(0)}} \langle \psi | \tilde{T} | \psi \rangle_g \right) \tag{S24}$$

here $\bar{T}$ is the average temperature, $e_{Si}^{TO} = dn_{Si}/dT = 1.8 \times 10^{-4}$/K is the TO coefficient of silicon (Si) material in PhCW, $-\mu p / T^2$ is the TO coefficient of gas, the subscripts Si and g represent the spatial overlapping integration $\int_{Si,g} \tilde{T} \psi^2 dV$ in the waveguide region and gas region, respectively. According to Eq. (S24), the phase modulation induced by Si and gas has opposite signs, and the overall phase modulation of the probe beam is amplified by the group index $n_g$.

### 1.4 Dual slow-light enhanced PTS

To illustrate the dual slow-light enhanced PTS, we define a normalized PT efficiency, $k^*$, representing that the pump power transfers to the phase change of the probe modes, written as:

$$k^* := \frac{\Delta\tilde{\phi}}{\alpha C L P_p} \tag{S25}$$

We numerically calculate $k^*$ using Eqs. (S14), (S19), (S24) and (S25) based on finite element method (FEM). Fig. S1(a) shows the calculated results for the two terms in Eq. (S24), The PT phase modulation in Si-based PhCW region is more than two orders of magnitude larger than that in gas region, which is mainly due to the large TO coefficient of Si compared to that of $C_2H_2$ gas ($-0.91 \times 10^{-6}$ /K).

Based on the calculated frequency response, it is sufficient to consider only the overlapping integration in the PhCW region $\langle\psi|\tilde{T}|\psi\rangle_{Si}$ and neglect the temperature change in the gas region. The solution of Eq. (S19) and (S20) may be approximated in the form of:

$$\tilde{T} = -\frac{A_{eq}}{\kappa_{Si} + i\omega_m \rho_{Si} c_{Si} A_{eq}} Q \propto \frac{1}{\kappa_{Si}} \frac{n_g}{n_{gas}^{(0)}} \alpha C \gamma I^{(0)} \tag{S26}$$

here $A_{eq}$ is the equivalent heat conduction area of the temperature field in the PhCW domain, and we have assumed that the modulation frequency $\omega_m \ll 2\pi f_{3dB}$ so that $\tilde{T} \propto 1/\kappa_g$. The value of $A_{eq}$ is determined to be 74.7 $\mu m^2$, which can be derived from the frequency response curve. The approximation given by Eq. (S26) is valid if the numerical simulation provides a frequency response that can be fitted by the Lorentz function. Fig. S1(b) shows the frequency response approximated using Eq. (S26), where the relative error is only 0.6% for $\omega_m \ll 2\pi f_{3dB}$. Combining Eq. (S26) with (S24), we finally get the expression for the PT phase modulation as:

$$\Delta\tilde{\phi} = k^* \alpha C L P_p \propto \frac{2\pi n_{gp} n_{gb} L}{\lambda_b \kappa_{Si}} e_{Si}^{TO} \alpha C \gamma P_p \tag{S27}$$

here $P_p$ is the pump optical power. Eq. (S27) shows the dual slow-light enhancement of PT phase modulation, $n_{gp}$ and $n_{gb}$ denotes the group indices at pump and probe wavelengths, respectively.

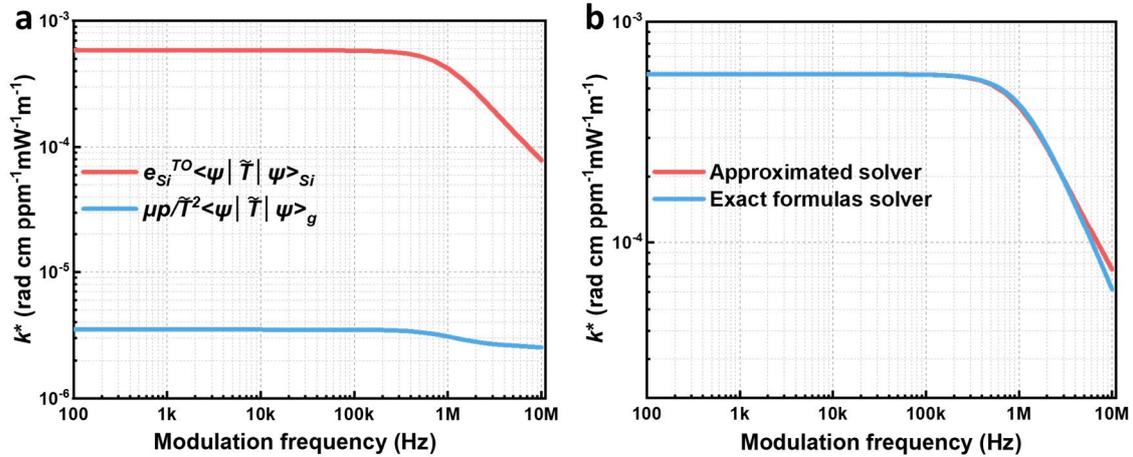

**Fig. S1 Calculation of the $k^*$ coefficient. (a)** Calculated $k^*$ value due to the PT effect of Si and gas. **(b)** Comparison of the $k^*$ value using the approximated and exact formulas solver of heat conduction.

## S2. Design of the dual slow-light PhCW

*2.1 Optimization of the PhCW structure*

The three-dimensional (3D) model of the suspended two-dimensional (2D)-PhCW is depicted in Fig. S2(a). Here several PhCW parameters are defined: $a$ is the lattice constant, $R$, $r_0$, and $r_1$ are the radius of three kinds of defect holes, and $h$ is the height of the air-based bottom cladding. To facilitate slow light in the near-infrared waveband, it is critical to maintain a sufficiently wide bandgap in the photonic crystal. Fig. S2(b) demonstrates the bandgap range varying with the ratio $R/a$. As $R/a$ increases, the normalized frequency range of the bandgap widens, and the central frequency correspondingly ascends. At a normalized frequency ($a/\lambda$) of 0.5, the existence of light cone results in the narrowing of the bandwidth in the guided mode curve, turning some modes into leaky modes. After evaluating both the bandgap range and the central frequency, an optimal $R/a = 0.38$ is selected. As shown in Fig. 2(a) of the main manuscript, two modes are generated in the photonic bandgap, which can be divided into even and odd modes based on the symmetry of the $z$-component of the magnetic field along the waveguide center plane. Since the odd mode can only be excited by the higher-order mode of the strip waveguide, the even mode, which can be easily excited by the fundamental mode, is excited in this PhCW structure.[7] To enhance gas absorption, the PhCW parameters are optimized to achieve a large spatial confinement of the energy density $\gamma$. Fig. S2(c) reveals the relationship of $\gamma$ and guided even mode bandwidth ($B_w$) with $r_0/R$. When $r_0 \leq 0.3R$, the curve of the even mode coincides with the flat mode near the band edge, rendering the slow-light mode in this overlapping region unexcitable. Consequently, only the bandwidth of the even mode is calculated for $r_0 \geq 0.3R$. As $r_0$ increases, $\gamma$ and $B_w$ show inverse trends, and an optimal $r_0$ of $0.6R$ is selected in this design. Fig. S2(d) shows the variations of $\gamma$ and $B_w$ with $r_1/R$ for $r_0 = 0.6R$, and $r_1 = 0.7R$ is chosen as the optimized value. The optimal parameters of the 2D-PhCW are as follows: $a = 500$ nm, $R/a = 0.38$, $r_0 = 0.6R$, $r_1 = 0.7R$, and $h = 2$ μm.

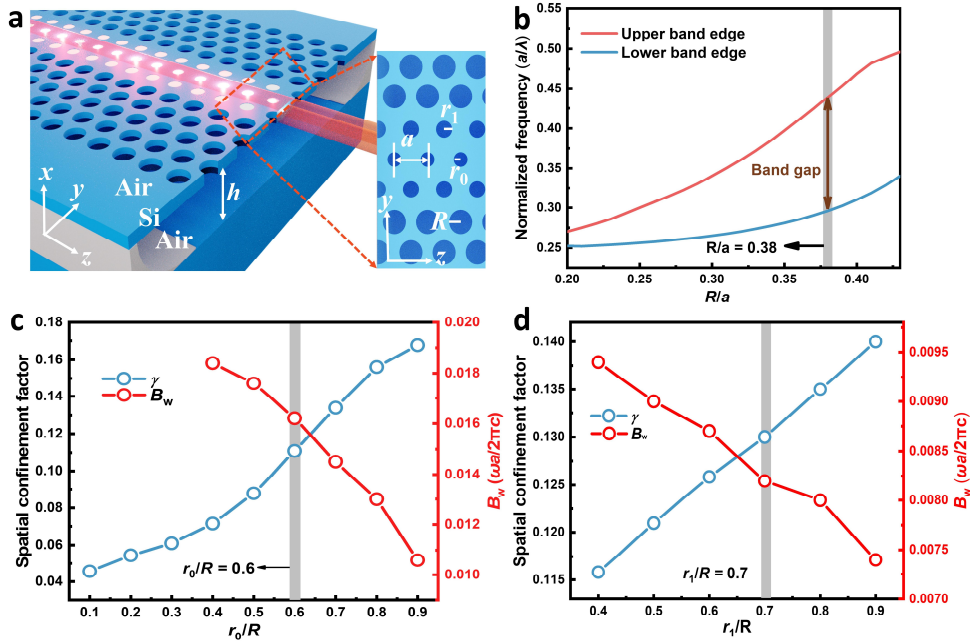

**Fig. S2 Optimization of the PhCW structure. (a)** Structure of the suspended 2D-PhCW and corresponding dimensional parameters. Right inset shows the $y$-$z$ plane with parameters of lattice constant ($a$), the radius of three kinds of defect holes ($R$, $r_0$, and $r_1$), and the height of the air-

based bottom cladding ($h$). **(b)** Photonic band gap *versus* $R/a$. The spatial confinement of the energy density $\gamma$ and guided mode bandwidth ($B_w$) *versus* **(c)** $r_0/R$ for $r_1=0$, and **(d)** $r_1/R$ for $r_0 = 0.6R$.

## 2.2 Design of the subwavelength grating coupler (SWGC) structure

The schematic structure of the SWGC is shown in Fig. S3(a). Fig. S3(b) shows the scanning electron microscopy (SEM) image of the fabricated SWGC with parameters of $P_1$=800 nm, $L_1$=400 nm, $P_2$=400 nm, $L_2$=80 nm. Here $P_1$ refers to the period of the grating coupler, $P_2$ is the period of the etched trench array, $L_1$ and $L_2$ are the length and width of the trench, respectively. The SWGC output is tapered into a 100-μm-long, 500-nm-wide strip waveguide. Given the polarization sensitivity of the SWGC, it exhibits high coupling efficiency specifically for TE-polarized light.[8] Therefore, to measure the coupling loss of the SWGC accurately, the input light is first adjusted to the TE-polarized state using a polarization controller, aligning with the maximum reading on the optical spectrum analyzer. The optimal orientation for minimizing coupling loss is achieved when the input/output optical fiber is inclined at an angle of 15°. The resultant coupling loss of the total input/output SWGC is shown in Fig. S3(c), indicating a total two-facet loss of 18 dB at ~1531 nm (as pump wavelength, *i.e.*, ~9 dB/facet) and 17 dB at ~1572 nm (as probe wavelength).

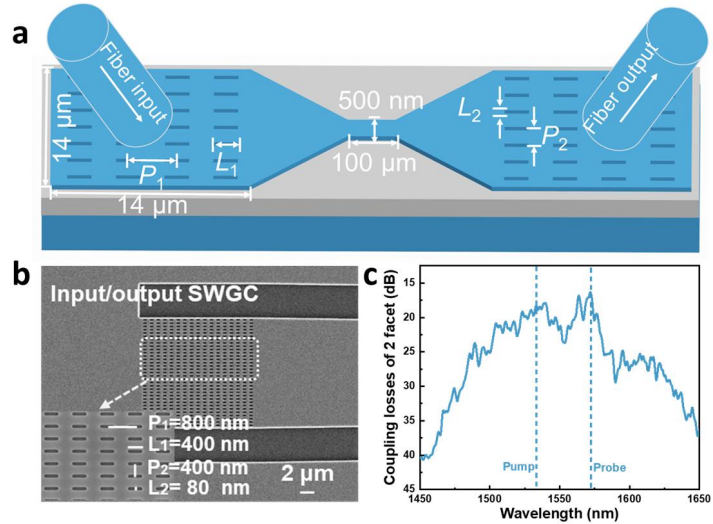

**Fig. S3 Design of the SWGC structure. (a)** Structure of the SWGC with optical fiber as input/output. Several structural parameters of the SWGC and strip waveguide are labeled for illustration **(b)** SEM image of the SWGC with parameters of $P_1$ = 800 nm, $L_1$ = 400 nm, $P_2$ = 400 nm, $L_2$ = 80 nm. **(c)** Measured two-facet coupling loss with input/output SWGCs.

## 2.3 Design of the taper waveguide structure

We design a taper waveguide to achieve a single $TE_0$ mode transmission to eliminate the influence of high-order modes (such as $TE_1$) in MZI. Fig. S4(a) shows the variation of effective refractive index with different waveguide widths for $TE_0$ and $TE_1$ modes. Inset shows a partial SEM image of the taper structure. The single-mode waveguide is designed to be 500 nm in width, where the waveguide only supports $TE_0$, while $TE_1$ and higher-order modes are cut off. As shown in Fig. S4(b), we further simulate the propagation of the $TE_0$ and $TE_1$ modes along the taper region using the 3D-FDTD method, which shows that the average transmission coefficient of the $TE_0$ mode is ~95%, while the $TE_1$ mode is only ~2%.

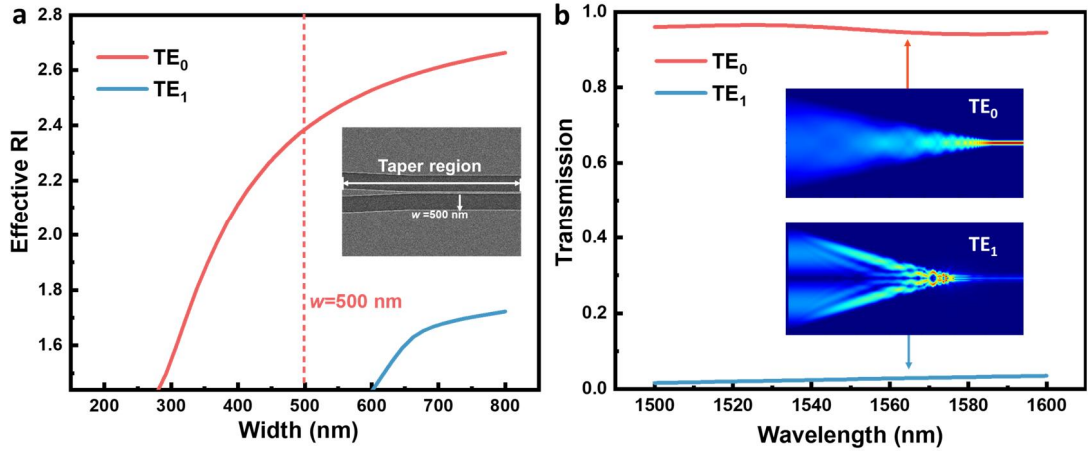

**Fig. S4 Design of the taper structure.** (a) Effective refractive index ($n_{\text{eff}}$) versus the strip waveguide width. Inset shows a partial SEM image of the taper structure with a narrowed width of 500 nm. (b) Simulated transmission of the $TE_0$ and $TE_1$ modes along the taper waveguide.

*2.4 Design of the transition structure*

Due to the significant difference in the group indices of the strip waveguide and the 2D-PhCW, the device may suffer high coupling loss from the strip waveguide to the PhCW.[9] Here we design a transition structure near the two ends of the PhCW to mitigate these losses, featuring a gradually changing hole radius of the central defect. The simulation models without/with transition structures are shown in Figs. S5(a) and (b), which utilize two power monitors ($IT_0$ and $OT_1$) to gauge the input/output powers.

As the radius $r_0$ of the central defect hole decreases, the guided mode frequency in the dispersion diagram drops, concurrently lowering the group index at that frequency. By finely adjusting $r_0$, a smooth transition in the group index can be achieved. The transmission spectrum with/without the transition structure is simulated using Ansys Lumerical FDTD software. As shown in Fig. S5(c), the resultant transmission exhibits a flatter profile across the interested slow-light region compared to that of the non-transition design.

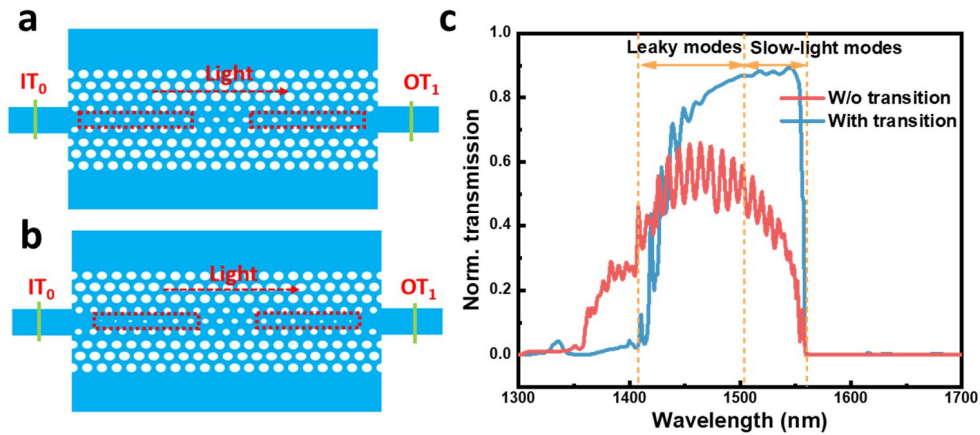

**Fig. S5 Design of the transition structure near the two ends of the PhCW.** (a) Without and (b) with transition design for illustration. (c) Simulated transmission spectrum of the TE even mode with and without transition structure.

## S3. Device fabrication process

Electron Beam Lithography (EBL) and Inductively Coupled Plasma (ICP) etching processes are first used to fabricate the on-chip devices, including input/output SWGCs, photonic crystal patterns, and other strip waveguides in the MZI structure on the top-Si layer of the Silicon-on-Insulator (SOI) wafer. The subsequent step involves wet-etching in the SOI interlayer using a $SiO_2$ buffer oxide etching (BOE) solution, which is a mixture of 40% $NH_4F$ and 49% HF solution with a volume ratio of 6:1. A suspended 2D-PhCW structure is obtained with air serving as both the top and bottom cladding layers. However, during the etching process, BOE may erode the $SiO_2$ that supports both ends of the PhCW, leading to the collapse of the strip waveguide. To address this issue, we first spin-coat the non-suspended area with photoresist and only wet-etch the photonic crystal area with BOE solution. Detailed fabrication process is referred to Fig. S6.

Step 1: Pre-process the SOI wafer with a 220-nm-thick top Si layer and a 2-μm-thick $SiO_2$ layer.

Step 2: ZEP 520A EB photoresist is spin-coated on the wafer with a thickness of ~300 nm.

Step 3: EBL is employed to fabricate the first-layer SWGCs, photonic crystal patterns, and other strip waveguides in the MZI structure.

Step 4: ICP is used to etch patterns on the Si layer. The thickness of the top Si and the diameter of the 2D-PhCW air hole determine the etching speed and specific dry etching time. The difficulty in discharging chemical reaction products is closely related to the diameter of the air hole. The smaller the air hole, the more difficult it is to remove, and the corresponding etching rate is slower. The etching in this step must be thorough, and the air hole needs to be completely etched to ensure that the subsequent wet etching can proceed smoothly, as the BOE solution infiltrates from the small hole.

Step 5: After removing the photoresist and ultrasonic cleaning, an asymmetric 2D-PhCW structure is obtained with air as the top cladding and $SiO_2$ as the bottom cladding.

Step 6: AZ701 ultraviolet (UV) photoresist is spin-coated.

Step 7: UV lithography. The UV photoresist in the photonic crystal area is removed, while the UV photoresist in other areas is retained to prevent BOE solution from corroding other areas.

Step 8: Immerse the wafer into the BOE for about 3 minutes until the $SiO_2$ under the PhCW is completely etched. A suspended 2D-PhCW is obtained with air serving as both the top and bottom cladding layers.

Step 9: To remove the UV photoresist, the wafer is first rinsed with deionized water, and then soaked in the photoresist removal solution for about 15 minutes.

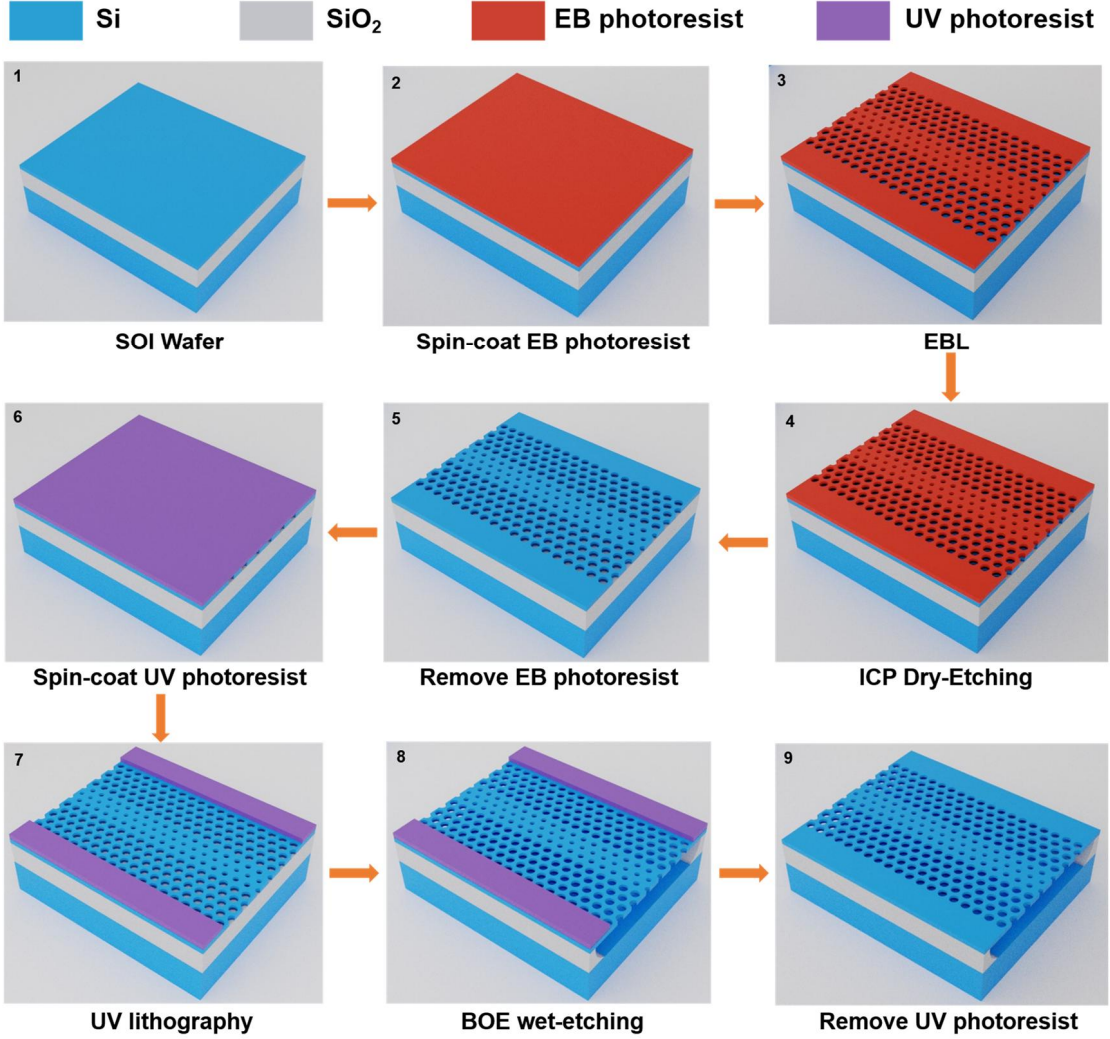

**Fig. S6 Fabrication process of the suspended 2D-PhCW.** The whole MZI structure is fabricated in-one-go on the same SOI wafer. EBL: electron beam lithography, UV lithography: ultraviolet lithography, BOE: buffer oxide etching.

## S4. Analysis of the PhCW-based MZI

Since the loss of the slow-light PhCW increases with $n_g$ value, there is a trade-off in the MZI configuration, as increased probe $n_g$ value enhances $\Delta\phi$, but the excessive loss induced by high $n_g$ could adversely affect the fringe contrast $v$ and overall PTS signal. Fig. S7 illustrates an on-chip MZI for the probe beam. A narrow-linewidth probe laser source is launched and divided into two beams through a Y-splitter. The two beams pass through the sensing and reference arms, respectively, with the PhCW at the sensing arm. Then the beams are re-combined at a Y-combiner and interfered on a photodetector (PD), which generates the output signal $I_{\text{out}}(t)$. The electrical field in the sensing arm can be expressed in the form of $E_s(t) = E_{0,s} \exp[j(wt + \phi_s + \Delta\phi(t))]$, while the electrical field in the reference arm expressed as $E_r(t) = E_{0,r} \exp[j(wt + \phi_r)]$. Here $E_{0,s}$ and $E_{0,r}$ represents the electrical strength in the sensing and the reference arms, respectively. $w$ the angular frequency of the optical wave, $\phi_s$ and $\phi_r$ the initial phase or static phase factor of the sensing and reference arms, respectively. $\Delta\phi(t)$ the time-varying PT-induced phase modulation. The output of the PD after the MZI can be described as

$$I_{\text{out}}(t) = E_{0,s}^2 + E_{0,r}^2 + 2E_{0,s}E_{0,r}\cos[\phi_s - \phi_r + \Delta\phi(t)] \quad (S28)$$

Eq. (S28) includes a direct current (DC) term and an alternating current (AC) term. The AC term is related to the time-dependent $\Delta\phi(t)$. In this setup, if the probe wavelength is tuned to operate at a quadrature (Q) point of the MZI, that is $\phi_s - \phi_r = n\pi + \pi/2, (n = 0,1,..)$ and assuming weak absorption $\Delta\phi(t) \ll 1$. The MZI output voltage $V(t)$ can be simplified as

$$V(t) = \eta_{PD}R_T\left[P_s + P_r - 2\sqrt{P_sP_r}\,\Delta\phi(t)\right] = (V_s + V_r)[1 - v\,\Delta\phi(t)] \quad (S29)$$

where $\eta_{PD}$ the PD responsivity, $R_T$ the transimpedance, $v$ the fringe contrast, written as:

$$v = \frac{V_{max} - V_{min}}{V_{max} + V_{min}} = \frac{2\sqrt{V_sV_r}}{V_s + V_r} \quad (S30)$$

The PTS signal amplitude ($S$) is defined as the 1$f$-demodulated voltage amplitude, which is proportional to the average probe power ($P_{avg}$) at the PD, $v$, and $\Delta\phi$, expressed as

$$S \propto P_{avg}\, v\, \Delta\phi \quad (S31)$$

The accumulated PT-induced phase modulation $\Delta\phi$ could be obtained from Eq. (S27), which generally increases with $n_{gb}$. However, with the increase of $n_{gb}$, the contrast $v$ decreases as the $n_g$-induced insertion loss $\alpha_L$ increases, resulting in the degradation of $S$. Therefore, there is a trade-off between $\Delta\phi$ and $v$ with an optimal $n_{gb}$ value. According to Eq. (S31), $S$ is proportional to the product of $\Delta\phi$ and $v$. The results of $\Delta\phi$, $v$ and $S$ in terms of $n_{gb}$ is presented in Fig. 4(d) of the main manuscript.

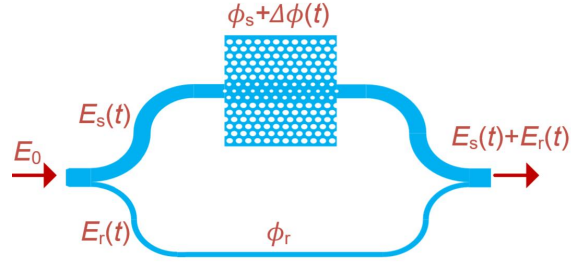

**Fig. S7 On-chip MZI model for the probe beam.** PhCW is in the sensing arm and a rectangular waveguide serves as the reference arm.

## S5. Evaluation of the dual slow-light enhancement factor

To evaluate the dual slow-light enhancement factor, we performed two series of experiments to obtain the relationship between the enhanced phase modulation and the product of $n_{gp}$ and $n_{gb}$. In the first series, we measured the PTS signals with a fixed pump wavelength (1531.58 nm, $n_{gp}$=15) and varied 22 probe wavelengths from 1540 nm to 1573 nm. In the second series, we measured the PTS signals with a fixed probe wavelength (1572 nm, $n_{gb}$=81) and varied 14 pump wavelengths from 1520 nm to 1540 nm aligned to the $C_2H_2$ absorption lines. Since the optical power and contrast can be normalized based on Eq. (S31), for clarity, the relationships of the three series (fixed $n_{gp}$ and varied $n_{gb}$, fixed $n_{gb}$ and varied $n_{gp}$, varied $n_{gp} \times n_{gb}$) are shown separately in Figs. S8(a)-(c). Three linear relationships of phase modulation (normalized value against its maximum) on $n_{gp}$, $n_{gb}$, and $n_{gp} \times n_{gb}$ are obtained with $R^2 > 0.94$. The measured and fitted relationships of Fig. S8(c) is shown in Fig. 4(e) of the main manuscript.

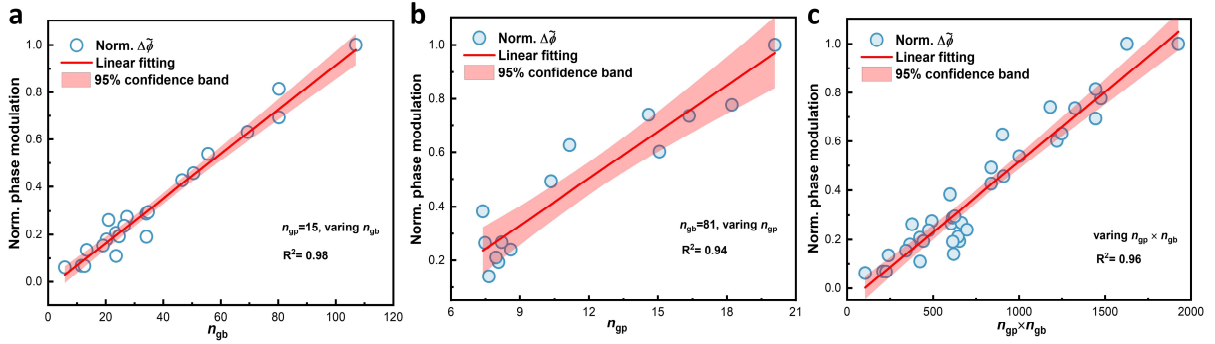

**Fig. S8** Measured $\Delta\tilde{\phi}$ (normalized against its maxima) as a function of $n_{gp}$ or $n_{gb}$. **(a)** Fixed $n_{gp}$ with varying $n_{gb}$. **(b)** Fixed $n_{gb}$ with varying $n_{gp}$. **(c)** Varying $n_{gp} \times n_{gb}$.

### S6. Evaluation of the stability of the on-chip MZI for PTS measurement

We compared the stability of the integrated on-chip MZI with that of a fiber-optic MZI with a sensing PhCW placed in sensing arm of the MZI and an external single-mode fiber as reference one. As shown in Fig. S9(a), the fiber-based MZI is actively stabilized at Q-point with a standard deviation (std.) of 0.45° using bulky servo-feedback equipment. The on-chip MZI can stably operate at Q-point with std. of 0.5°, comparable to that of the servo-locked fiber-based MZI. Notably, this stability is achieved without any electronic servo-loop equipment to stabilize the interferometer at the Q-point. In this manner, compared to the fiber-optic MZI with the benchtop locking circuit equipment, the on-chip MZI markedly reduces system complexity and minimizes the footprint to ~0.6 mm², achieving a dramatic dimension reduction while still maintaining high stability even without any servo-locking equipment. Furthermore, as shown in Figs. S9(b)-(d), we integrate two optical fiber arrays (OFAs) with the on-chip MZI to package the device into a fiber-coupled chip and then seal the chip into a compact gas cell for gas sensing.

To operate the device at the Q-point of the on-chip MZI, we select the mid-point of the wavelength-swept probe cycle, which corresponds to the average transmission level (~0.401 V) of the MZI fringe, as shown in Fig. S10(a). To maintain stable quadrature operation, we implemented thermal stabilization of the entire gas cell, including the MZI and other waveguide structures within it. As shown in Fig. S10(b), we performed an experiment in which the temperature of the thermal controller was adjusted from 24.97 °C to 25.02 °C in increments of 10 mK. We then recorded the variation in the Q-point (i.e. phase) change, which enables us to obtain the temperature-to-phase conversion coefficient of 191°/K. In this work, we control the temperature of the gas cell to ±2.6 mK, which ensures the stable operation (standard deviation of 0.5°) around the Q-point.

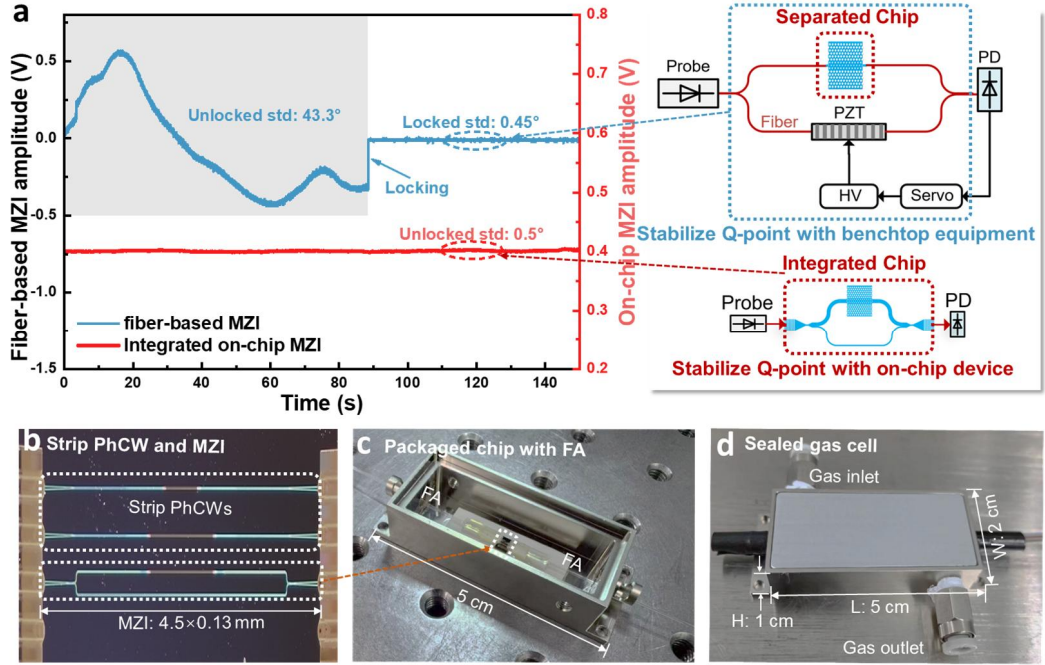

**Fig. S9 Evaluation of the on-chip MZI stability. (a)** Comparison of the stability and dimension between the off-chip (fiber-based) MZI with servo-control and on-chip integrated MZI without the need for servo control. PZT: Piezoelectric transducer, HV: High voltage amplifier, Servo: Servo controller for phase-locking. Separated chip: Fiber-optic MZI. Integrated chip: Integrated on-chip MZI. **(b)** Fabricated PhCW-based strip waveguide and on-chip MZI with optical fiber arrays (OFAs) fixing at two ends. **(c)** Packaged fiber-coupled chip with two OFAs, which are integrated with input/output SWGCs at an optimized tilt angle of 15°. **(d)** Sealed gas cell with a compact dimension of $5\times2\times1$ cm$^3$ and an inlet/outlet for gas flow.

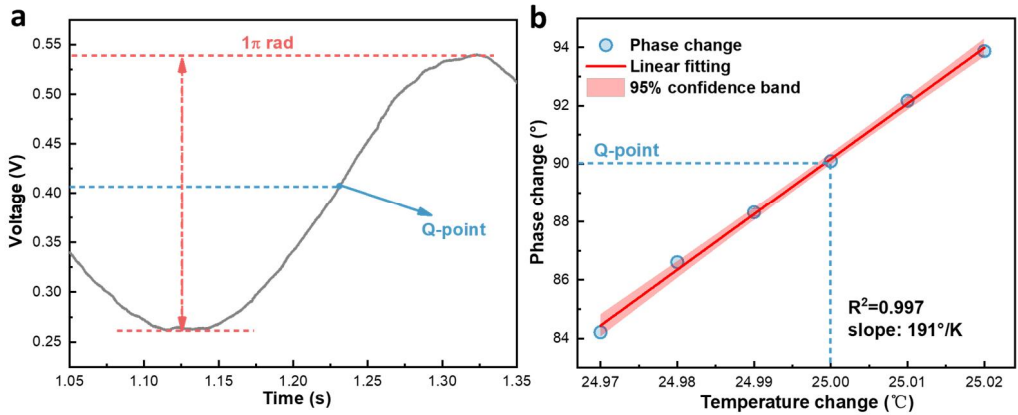

**Fig. S10 Evaluation of the temperature stabilization. (a)** Sweeping the probe wavelength around Q-point. **(b)** Relationship between the phase variation and temperature change using thermal controller.

## S7. Evaluation of the dual slow-light PTS noise level

We conducted experiments to assess the noise level of the dual-slow-light PTS, which is characterized by the measured probe power spectral density (PSD) on the PD output using an electrical spectrum analyzer (ESA). As shown in Fig. S11(a), we initially tested the scenario where the probe beam is directly detected (pump off, without passing through the MZI), the PSD shows the relative intensity noise (RIN) of the probe laser, as represented by the red line in Fig. S11(b). Subsequently, when both the pump and probe beams were coupled into the MZI in a 1% $C_2H_2$ environment, a PTS signal at 50 kHz was observed (blue line in Fig. S11(b)). For comparison, the PTS signal in pure $N_2$ environment is shown as the purple line in Fig. S11(b). We would like to clarify that the observed PSD values for $C_2H_2$ (blue line) and $N_2$ (purple line) at 50 kHz represent the signal data, not the noise level.

It is important to note that the purple signal under the $N_2$ environment serves as a baseline, which remains constant under fixed pump power and is independent of gas concentrations. This phenomenon is attributed to the Kerr effect in silicon. The 100× signal increase in the presence of $C_2H_2$ compared to $N_2$ indicates that the PTS signal for 1% $C_2H_2$ is 100 times larger than the baseline value of $N_2$, as demonstrated in Fig. S11(c). Since the baseline does not depend on gas absorption, it can be removed by measuring the PTS signal with the pump wavelength periodically scanning across the absorption line. The background traces around these signals correspond to the noise levels, which, for the above three scenarios, are close to the probe RIN of $3\times10^{-7}$ V/√Hz at 50 kHz, indicating that the on-chip MZI and the pump laser contribute minimally to the overall system noise.

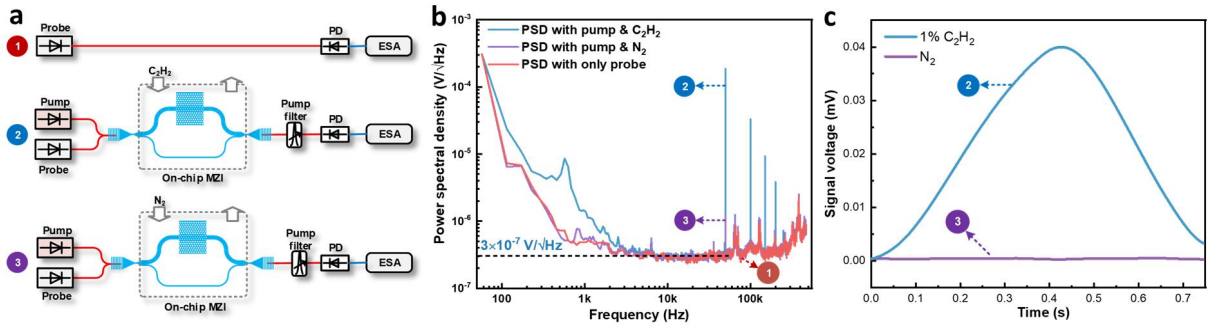

**Fig. S11 Evaluation of the PTS noise. (a)** Schematic of the three PSD measurement set-ups. **(b)** The measured PSD without using the pump and with probe beam directly detected by PD (red line, case 1); pump and probe beams are coupled into the MZI placed in 1% $C_2H_2$ environment (blue line, case 2); pump and probe are coupled into the MZI in $N_2$ environment (purple line, case 3). **(c)** The detected $1f$ signal for 1% $C_2H_2$ (blue line, case 2) and for pure $N_2$ (purple line, case 3) at 50 kHz when the pump wavelength is scanned across a $C_2H_2$ absorption line.

## S8. Comparison of the $k^*$ for different fibers/waveguides

To compare the PT efficiency for different waveguide structures, we apply FEM to calculate their $k^*$ value with 2D models (*i.e.,* invariant along the $z$-axis). Since the cross-section of PhCW is not uniform along the $z$-axis, the PhCW-based PTS is simulated in a 3D model. The PhCW is modeled with 25 units of photonic crystal lattices along the $z$-axis, 14 units along the $x$-axis, and only one layer on the $y$-axis. An additional Si layer is placed at the bottom of the PhCW region with a distance of 2 μm, which acts as the substrate of the suspended PhCW. For each 2D model, the pump mode analysis is first conducted and is used as an input for the heat source. The heating process is calculated using Fourier's heat transfer equation with the outer boundary set as a constant temperature of 293.15 K. As for the 3D model of the PhCW, the input and output ports are set to be "Numeric" and are used for simulating the pump mode distribution. The insertion loss is excluded in evaluating its $k^*$ value through normalization. For thermal fields, a symmetry boundary condition is applied to the optical input and output surfaces to simulate a scenario where the structure is infinitely periodic along the z-axis. The other boundaries surrounding the PhCW are allowed to transfer the heat to the air via natural convective heat flux.

Fig. S12 shows the numerical simulation and some experimental results. The PhCWs exhibit $k^*$ value of $\sim 5.8 \times 10^{-4}$ rad·cm·ppm$^{-1}$·mW$^{-1}$·m$^{-1}$ (<1 MHz), while HCF-based $k^*$ values are at the level of $\sim 10^{-6}$ rad·cm·ppm$^{-1}$·mW$^{-1}$·m$^{-1}$ (<10 kHz), and strip waveguide-based $k^*$ values are at the $\sim 10^{-5}$ rad·cm·ppm$^{-1}$·mW$^{-1}$·m$^{-1}$ level (<10 kHz). It is observed that the PhCWs demonstrate 2−3 orders of magnitude higher than HCFs, and 1−2 orders of magnitude higher than conventional strip waveguides. The primary reasons for this advantage include the smaller mode field area, the larger TO coefficient of Si, and most importantly, the dual slow-light enhancement effect of the PhCW. Besides, the 3-dB bandwidth of PhCW is the widest, which is mainly due to the heat dissipation mechanism of Si with high thermal conductivity. Such a fast heat conduction would limit the heat from accumulating but rather dissipating, which lowers the maximum PT efficiency of the PhCW but broadens the frequency response. According to Eq. (S27), there is a trade-off between the maximum PT efficiency and its bandwidth:

$$\begin{cases} \tilde{T}_{\max} = \dfrac{A_{\mathrm{eq}}}{\kappa_{\mathrm{Si}}} \tilde{Q} \\ \omega_{\mathrm{3dB}} = \dfrac{\kappa_{\mathrm{Si}}}{\rho c A_{\mathrm{eq}}} \end{cases} \quad (S32)$$

here the heat conduction is described by $A_{\mathrm{eq}}/\kappa_{\mathrm{Si}}$. As described by Eq. (S32), for a same fiber/waveguide structure, a larger 3-dB bandwidth means a lower $T$ and hence $k^*$. Compared to the other structures, the PhCW shows both significantly larger $k^*$ and 3-dB bandwidth, which is mainly attributed to the dual slow-light enhancement regime.

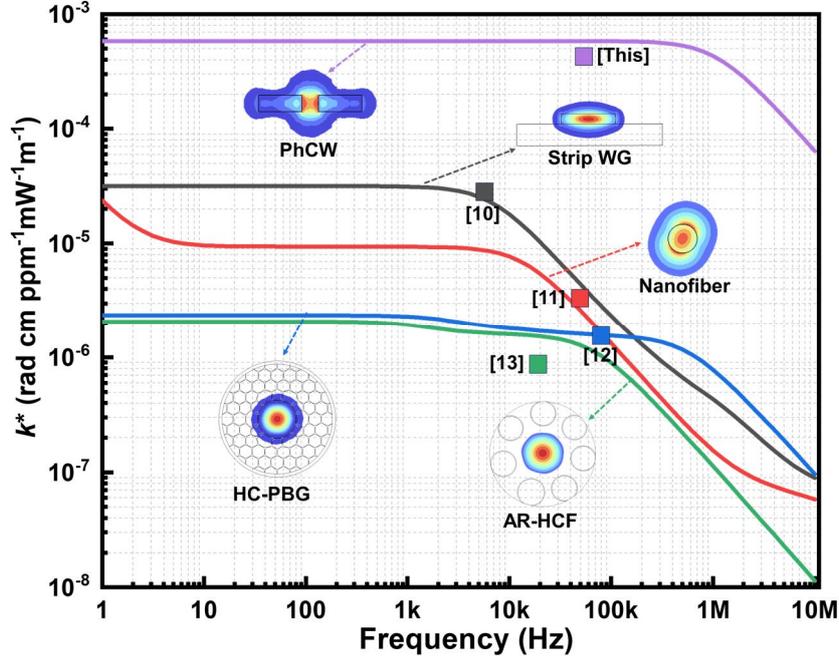

**Fig. S12 Comparison of the $k^*$ coefficient across different fibers and waveguides.** Inset images show the cross-sectional optical fields of different fibers and waveguides. Solid square points represent the corresponding experimental data. $k^*$ of PhCW in this work is measured to be $3.6\times10^{-4}$ rad·cm·ppm$^{-1}$·mW$^{-1}$·m$^{-1}$ @50 kHz. Strip WG: SOI-based rectangular waveguide ($3.05\times10^{-5}$ rad·cm·ppm$^{-1}$·mW$^{-1}$·m$^{-1}$ @6 kHz).[10] Nanofiber: Nanoscale tapered single-mode fiber ($3.26\times10^{-6}$ rad·cm·ppm$^{-1}$·mW$^{-1}$·m$^{-1}$ @62 kHz).[11] HC-PBG: Hollow-core photonic bandgap

fiber (1.12×10⁻⁶ rad·cm·ppm⁻¹·mW⁻¹·m⁻¹ @100 kHz).[12] AR-HCF: Anti-resonant hollow-core fiber (8×10⁻⁷ rad·cm·ppm⁻¹·mW⁻¹·m⁻¹ @19 kHz).[13]

## S9. Comparison with the single slow-light PhCW for DAS measurement

Before employing the single slow-light PhCW for direct absorption spectroscopy (DAS) measurement, we first measure the transmission and insertion loss of the PhCW. The schematic setup is depicted in Fig. S13(a). A broadband laser source (SC-5, YSL Photonics) is used as the input, which passes through a polarization controller to ensure a TE-polarized state. The broadband light is then introduced into the PhCW via fiber-SWGC coupling with an incident angle of 15°. The output is connected with another SWGC and then propagates into an optical spectrum analyzer (OSA, AQ6370D) with a resolution of 20 pm. The upper panel in Fig. 13(b) shows the transmission spectra of only two input/output SWGCs without PhCW (defined as I), two SWGCs with 1-mm-long PhCW (II) and 3-mm-long PhCW (III), respectively.

As shown in the lower panel of Fig. S13(b), the transmission loss is obtained by subtracting III from II, leading to an average value of ~5 dB/mm. The insertion loss of the PhCW is obtained by subtracting II from I, which includes the transmission loss of 1 mm PhCW, and the coupling loss of the strip waveguide-PhCW, excluding the input/output SWGCs. An increasing trend of the insertion loss as a function of wavelength is recorded, indicating a wavelength (*i.e.*, $n_g$)-related PhCW loss property.[14]

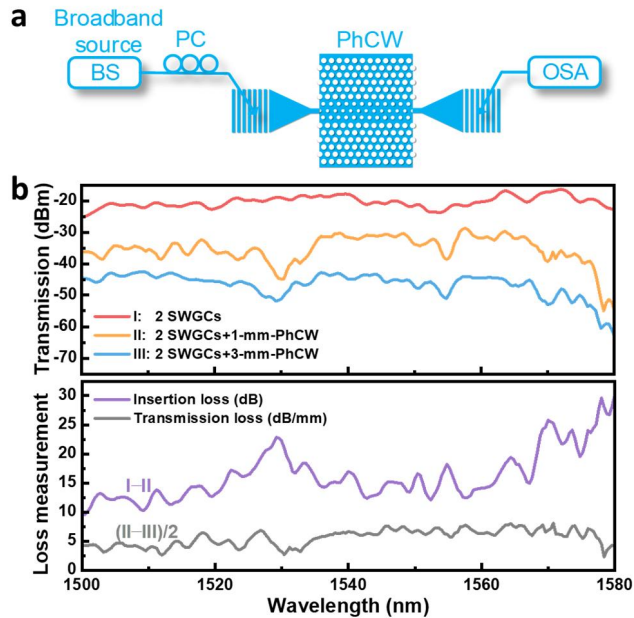

**Fig. S13 PhCW-based loss measurement.** **(a)** Schematic set-up. The PhCW is fabricated with two optical lengths of 1 mm and 3 mm. BS: Broadband source, PC: Polarization controller, OSA: Optical spectrum analyzer. **(b)** The upper panel shows the transmission spectra of two input/output SWGCs without PhCW, and two SWGCs with a 1-mm-long PhCW and a 3-mm-long PhCW. The Lower panel shows the transmission and insertion losses of the PhCW.

The schematic of the single slow-light PhCW-based DAS sensor is shown in Fig. S14(a). Unlike the dual-slow light PTS system, this setup only employs a pump laser so that the slow light effect on DAS can be investigated. The pump emission wavelength is tuned to 1531.58 nm by regulating the temperature controller and current controller, which is then coupled into the PhCW after passing through an erbium-doped fiber amplifier (EDFA). To mitigate low-frequency noise in the system, an acoustic-optic modulator (AOM) is used to modulate the intensity of the laser beam at a 50 kHz frequency. The beam is coupled to the PhCW via a single-mode fiber-based SWGC after passing through a polarization controller. The output beam from another SWGC is directly connected to a PD for the first harmonic (1$f$) signal demodulation. A sine wave at the same frequency generated by the signal generator is used as the reference signal of the lock-in amplifier (LIA). The 1$f$ signal demodulated by the LIA is recorded by a data acquisition card for further data processing. The structure and parameters of the DAS gas cell are the same as those of the PTS.

Fig. S14(b) displays the absorbance of the 2D-PhCW at four $C_2H_2$ concentrations. The red line representing a linear fit of the absorbance data demonstrates the good linearity of the DAS sensor. The group index can be calculated from the absorbance data, as the measured absorbance is expressed as $A_{exp.} = -\ln(I/I_0) = n_{gp}\alpha C\gamma L$, where $I$ and $I_0$ represent the signal amplitudes with $C_2H_2$ and $N_2$, respectively. The experimental $A_{exp.}$ is then compared with the theorical absorbance from the HITRAN database, expressed as $A_{theory} = \alpha C\gamma L$ with effective optical path length of $\gamma L$.[15] The $n_{gp}$ value at the pump wavelength of 1531.58 nm is determined to be 18. To evaluate the minimum detection limit (MDL) of the DAS sensor, Allan deviation analysis is performed based on the data collected in the $N_2$ environment, as shown in Fig. S14(c). The single slow-light PhCW DAS sensor achieves an MDL of 436 parts-per-million (ppm) with an average time of 45 s. Such sensitivity is about 36 times worse than that of dual slow-light PTS (12 ppm), which is mainly attributed to the absence of another slow-light $n_{gb}$ factor for enhancing phase modulation.

Another reason is the lower fringing noise of PTS over DAS.[10] As noted, fringing effect (i.e., etalon noise) is indeed a concern in integrated photonic systems due to reflections at material interfaces (e.g., Si/SiO$_2$ or Si/air) and waveguide line-edge roughness imperfections. In DAS, such fringes often dominate the noise baseline, particularly in low-concentration regimes where weak absorption signals can be obscured by periodic interference patterns. DAS typically uses a single laser for both excitation and detection, so any baseline fluctuation or fringing directly affects the signal. Moreover, increasing pump power in DAS amplifies both signal and fringe noise, limiting the achievable signal-to-noise ratio (SNR).

In contrast, PTS employs separate pump and probe beams, thus offers two inherent advantages that reduce sensitivity to fringing. (1) For the pump fringing, the pump excites the gas-induced local heating and a refractive index change. The resulting thermal perturbation is detected by the probe via interferometric phase modulations (e.g., through an MZI). Increasing pump power in PTS amplifies the signal without significantly increasing the noise baseline, since the noise is dominated by the probe beam. Only when the pump power becomes high enough that pump-intensity noise takes over, does the SNR cease to improve and fringe drift become problematic. In our experiments, even at pump powers of several hundred milliwatts, increasing the pump power continued to enhance the SNR, confirming that the noise is governed by the probe and that the system remains essentially insensitive to pump fringing. (2) As for the probe fringing, the probe is fixed in wavelength, which detects the modulated thermal signal due to absorption of intensity or wavelength modulated pump using with lock-in harmonic detection. This synchronous demodulation isolates the dynamic modulated signal from static and broadband background noise, including fringing. Any residual etalon effects in the probe path manifest as static patterns and are inherently filtered out by the lock-in amplifier. This unique characteristic makes PTS a background-free and ultralow-noise detection technique, rendering it insensitive to etalon noise.

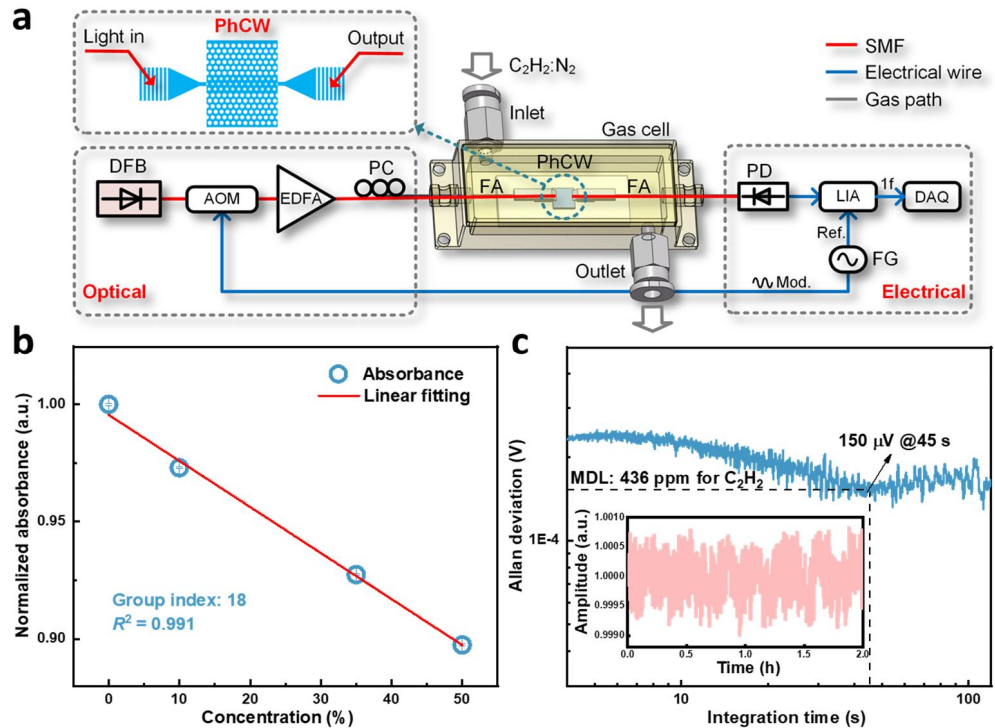

**Fig. S14 Single slow-light PhCW for DAS measurement. (a)** Experimental set-up. DFB: distributed feedback laser, AOM: acousto-optic modulator, EDFA: erbium-doped fiber amplifier, PC: polarization controller. LIA: lock-in amplifier, DAQ: data acquisition card. **(b)** Absorbance signal measurement with a pump wavelength of 1531.58 nm for $n_{gp}$ evaluation. Error bars of the absorbance show the standard deviation from five measurements. **(c)** Allan-Werle plot for MDL evaluation based on the inset long-term noise data.

## S10. Allan deviation analysis of the PhCW-PTS

In this experiment, the pump wavelength for PTS sensing was set to 1531.58 nm, corresponding to the P(11) $C_2H_2$ absorption line. For the Allan deviation characterization, we tuned the pump wavelength slightly away from the absorption line to 1531.27 nm, where the absorption of $C_2H_2$ is negligible. The group indexes at these two wavelengths are essentially the same. As shown in Fig. S15(a), two long-term noise measurements were conducted under different conditions. The blue line represents the noise in pure $N_2$ when the pump is adjusted to 1531.58 nm (on the absorption line), and the red line represents the noise in pure $N_2$ when the pump is adjusted to 1531.27 nm (off the absorption line). As shown in Fig. S15(b), Allan-Werle analysis is conducted to determine the optimal integration time and MDL under these two conditions. The results are averaged through a 1000-point window smoothing for accurate estimation, which show that at ~200 s averaging time, the Allan deviation reaches its 1σ MDL in both cases: 15 ppm (at 1531.58 nm) and 12 ppm (at 1531.27 nm), indicating minimal difference and confirming the reliability of the off-absorption line noise characterization. We further evaluated the signal fluctuation at the $C_2H_2$ concentration of 1000 ppm (at 1531.58 nm). As shown in Figs. S15(c) and (d), the 1σ fluctuation is estimated to be 18 ppm at an averaging time of 100 s, slightly larger than the 1σ MDL value obtained with pure $N_2$.

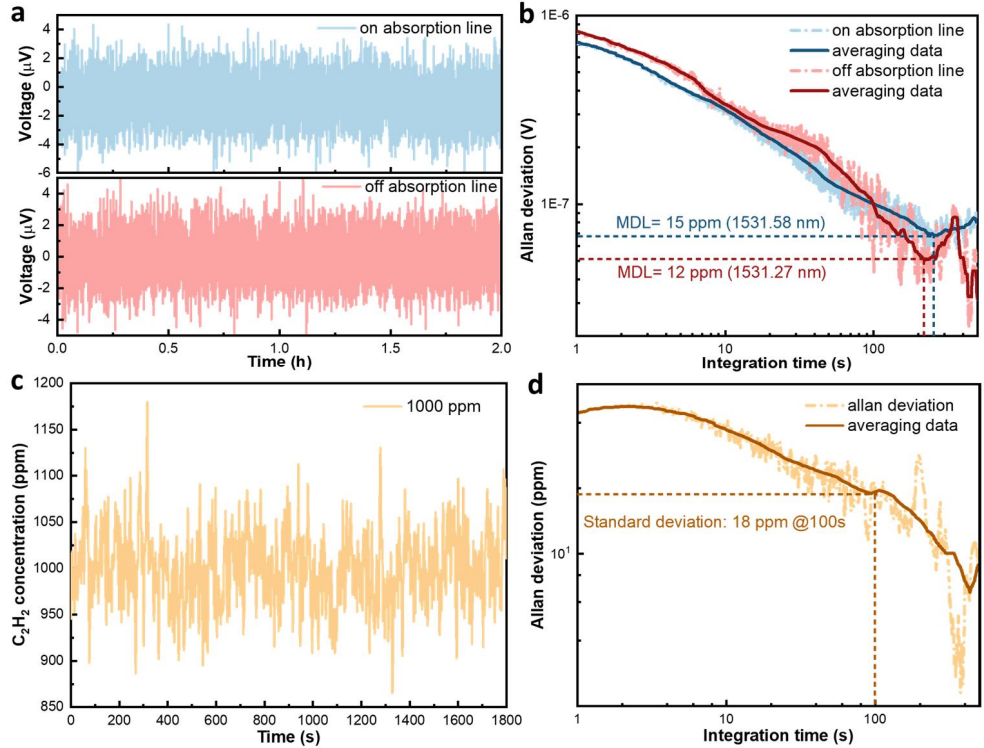

**Fig. S15 Allan deviation analysis for PTS measurement.** (a) Noise measurement in $N_2$ environment with the pump wavelength tuned on (blue lines) and off (red lines) the absorption line and (b) corresponding Allan-deviation analysis. (c) Signal fluctuation with a fixed $C_2H_2$ concentration of 1000 ppm and (d) corresponding Allan-deviation analysis.

## S11. Comparisons between different types of gas sensors

For comparison, we summarize the performance of the reported waveguide, free-space/fiber and slow-light gas sensors in Tables S1-S3. We use MDL, noise-equivalent absorption (NEA), NEA product length (NEA·L) and NNEA·L to evaluate the sensing performance. Here NNEA·L represents the power (P) and bandwidth (BW) normalized noise equivalent absorption coefficient and length product, expressed as NEA·L·P·BW$^{-1/2}$ with unit of W·Hz$^{-1/2}$.

### 11.1 Comparisons with photonic waveguide sensors

Waveguide absorption spectroscopic gas sensors reported so far include DAS and PTS. Compared to DAS, PTS-based approach offers several critical advantages. (1) DAS relies on detecting small absorption-induced intensity changes, which are highly susceptible to background fluctuations and fringing noise, particularly in low-concentration detection. In contrast, PTS measures thermally induced phase shifts via a modulated pump-probe scheme. This inherently rejects static background noise and enables ultralow-noise detection, significantly improving the SNR. (2) The sensitivity of PTS scales inversely with the optical mode area. Our integrated photonic platform features tight optical confinement and strong light–matter interaction, resulting in much higher sensitivity than traditional fiber or free-space optical systems. (3) PTS is well-suited to on-chip integration, allowing for compact, scalable, and robust sensor systems that outperform the bulky setups required for most DAS-based free-space/fiber approaches. (4) In Table S1, we provide a detailed comparison of our PTS sensor with state-of-the-art DAS sensors, demonstrating a 1–3 orders of magnitude improvement both in sensitivity and dynamic range. The calculation of NNEA·L requires the knowledge about the detection bandwidth. In Table S1, many of the references reporting the DAS waveguide sensors did not clearly state the bandwidth values. For those with known values of bandwidth, their NNEA·L values are in the range of $10^{-6}$ to $10^{-7}$ W·Hz$^{-1/2}$.[10,16] Our current PTS sensor achieves $1.8\times10^{-8}$ W·Hz$^{-1/2}$.

**Table S1** Comparison of the dual slow-light enhanced PTS sensor with waveguide gas sensors

| Year &ref. | Method | Gas λ(nm) | Waveguide type | Γ (%) | Loss (dB/cm) | L (cm) | BW (Hz) | P (mW) | MDL (time) | NEA·L | NNEA·L (W·Hz$^{-1/2}$) | Dynamic range |
|---|---|---|---|---|---|---|---|---|---|---|---|---|
| 2017[16] | DAS | $CH_4$ 1650 | Si-strip | 28.3 | 2 | 10 | 5.7 | 10 | 100 ppm (60s) | $4.6\times10^{-4}$ | $1.9\times10^{-6}$ | $>1.5\times10^2$ |
| 2020[17] | DAS | $CO_2$ 4240 | Si-strip# | 44 | 3 | 0.32 | Not stated | Not stated | 1000 ppm | $9.9\times10^{-2}$ | Not stated | $>1\times10^2$ |
| 2021[18] | DAS | $C_2H_2$ 2566 | $Ta_2O_5$-strip# | 107 | 6.8 | 2 | Not stated | 8.5 | 7 ppm (25s) | $3.6\times10^{-4}$ | Not stated | $1.4\times10^4$ |
| 2023[19] | DAS | $CH_4$ 3270 | Si-slot | 69 | 8.3 | 1.15 | Not stated | 1 | 0.3 ppm (50s) | $1.7\times10^{-5}$ | Not stated | $3.3\times10^3$ |
| 2024[20] | DAS | $CO_2$ 4230 | Ge-slot | 45.2 | 1.93 | 2 | Not stated | 1 | 0.627ppm (36s) | $4.6\times10^{-4}$ | Not stated | $>3.2\times10^2$ |
| 2024[10] | PTS DAS | $C_2H_2$ 1531 | ChG-strip | 10.5 | 2.5 | 2 | 0.94 | 6.3 | 4 ppm (170s) 65 ppm (100s) | $9.3\times10^{-6}$ $1.5\times10^{-4}$ | $6\times10^{-8}$ $9.6\times10^{-7}$ | $1.4\times10^5$ $1.5\times10^3$ |
| 2024[21] | PTS | $CO_2$ 2004 | LN-strip | 7.4 | 1.2 | 9.12 | 1.44 | 8 | 870 ppm (190s) | $1.1\times10^{-3}$ | $7.3\times10^{-6}$ | $>6.2\times10^2$ |
| This work | PTS | $C_2H_2$ 1531 | Si-2D PhCW | 234 | 50 | 0.1 | 0.94 | 12.5 | 12 ppm (200s) | $1.4\times10^{-6}$ | $1.8\times10^{-8}$ | $>2.5\times10^4$ |

*Γ*: effective confinement factor, defined as the product of $n_g$ and spatial confinement of the energy density *γ*. Si: Silicon, $Ta_2O_5$: Tantalum pentoxide, Ge: Germanium, ChG: Chalcogenide glass, LN: Lithium niobate, $CH_4$: Methane, $CO_2$: Carbon dioxide. In Refs. [10] and [21], fiber-optic (off-chip) interferometers are used for PTS measurement with complicated and bulky servo-locking equipment.[10,21] In this PTS setup, an on-chip MZI with an ultracompact footprint of 0.6 mm$^2$ is used without any external locking equipment but with comparable stability.

### *11.2 Comparisons with free space/optical fiber sensors*

We have included a detailed performance comparison in Table S2, highlighting the advantages of our on-chip sensor over existing free-space and fiber-optic sensors. Our on-chip PhCW-PTS sensor achieves an NEA·L of $1.4\times10^{-6}$, approaches the advanced free-space/fiber sensors, which typically require large, complex, and non-integrated setups. In terms of NNEA·L, free-space sensors based on cavities and multi-pass cells reach to the level of $10^{-4}$–$10^{-6}$ W·Hz$^{-1/2}$, while fiber-based PTS sensors have NNEA·L values ranging from $10^{-7}$ to $5.3\times10^{-9}$ W·Hz$^{-1/2}$. Our on-chip dual-slow-light enhanced PTS sensor achieves $1.8\times10^{-8}$ W·Hz$^{-1/2}$. Importantly, our sensor offers several distinct advantages beyond sensitivity: (1) The entire sensing device occupies only 0.6 mm$^2$, which is many orders of magnitude smaller than most free-space or fiber-based setups. (2) The silicon-based PhCW is fabricated using standard CMOS-compatible process, enabling low-cost mass production and seamless integration with electronic and photonic components. (3) Scalability for potential dense on-chip sensor arrays or deployment in portable, distributed, or wearable platforms where size, power and price are critical. (4) A compact, partly integrated architecture, including waveguide, interferometer, and gas cell, setting the stage for future inclusion of on-chip light sources and detectors.

**Table S2** Comparison of the on-chip sensor with free-space/optical fiber sensors

| Sensor types | Year& ref. | Methods | Gas & $\lambda$(nm) | Gas cell & length | $L_{eff}$ (m) | BW (Hz) | P (mW) | NEA·L | NNEA·L (W·Hz$^{-1/2}$) | Dynamic range |
|---|---|---|---|---|---|---|---|---|---|---|
| Free-space sensors | 2016[22] | TDLAS | CH$_4$ 3334 | 17 cm MPC | 54.6 | 10 | 10 | 7.1×10$^{-4}$ | 2.2×10$^{-6}$ | >1.2×10$^3$ |
| | 2016[23] | OF-CEAS | CH$_4$ 7390 | 80 cm cavity | 2760 | 0.97 | 180 | 1.5×10$^{-4}$ | 2.7×10$^{-5}$ | Not stated |
| | 2019[24] | OA-ICOS | CH$_4$ 1654 | 28 cm cavity | 27000 | 0.02 | 12 | 1.9×10$^{-3}$ | 1.6×10$^{-4}$ | >2.6×10$^2$ |
| | 2024[25] | CRDS | NH$_3$ 9556 | 50 cm cavity | 833 | Not stated | 40 | 4.5×10$^{-4}$ | Not stated | 2.3×10$^4$ |
| | 2024[26] | LITES | C$_2$H$_2$ 1530 | 5 cm MPC | 9.1 | Not stated | 1000 | Not stated | Not stated | 1.5×10$^4$ |
| Optical fiber sensors | 2021[27] | PTS-FPI | C$_2$H$_2$ 1530 | 5.5 cm HC-ARF | 0.05 | 0.094 | 125 | 1.3×10$^{-8}$ | 5.3×10$^{-9}$ | 6.2×10$^4$ |
| | 2015[12] | PTS-MZI | C$_2$H$_2$ 1530 | 10 m HC-PBF | 10 | 0.094 | 15.3 | 2.3×10$^{-6}$ | 1.2×10$^{-7}$ | 5.3×10$^5$ |
| On-chip sensor | This work | PTS-on chip MZI | C$_2$H$_2$ 1531 | 1 mm PhCW | 0.001 | 0.94 | 12.5 | 1.4×10$^{-6}$ | 1.8×10$^{-8}$ | >2.5×10$^4$ |

TDLAS: Tunable diode laser absorption spectroscopy, OF-CEAS: Optical feedback cavity-enhanced spectroscopy, OA-ICOS: Off-axis integrated cavity output spectroscopy, CRDS: Cavity ringdown spectroscopy, LITES: Light-induced thermoelectric spectroscopy. MPC: multi-pass cell, HC-PBF: hollow-core photonic bandgap fiber, HC-ARF: hollow-core anti-resonant fiber, MZI: Mach-Zehnder interferometer, FPI: Fabry-Pérot interferometer.

### *11.3 Comparisons with slow-light based sensors*

This work introduces a new dual slow-light scheme that departs fundamentally from previous PhCW/SWG gas sensors based on single slow-light effect. We confirm that the on-chip photothermal efficiency is proportional to the product of group indices of the pump and probe, which is attributed to $n_g$-amplified perturbation on imaginary and real part of the permittivity respectively. Leveraging this principle, we realize the first dual-slow light on-chip PTS, an inherently background-free and low-noise alternative to DAS. Dispersion engineering supplies the slow-light band to cover both pump absorption and probe phase modulation, resulting in a substantial sensitivity enhancement.

We provide a comprehensive comparison of current slow-light based gas sensors in Table S3, including PhCW and SWG sensors. Our dual slow-light PTS sensor demonstrates advantages in terms of NEA·L sensitivity and dynamic range, outperforming existing single slow-light waveguide sensors by 1–3 orders of magnitude. Furthermore, most previously reported PhCW/SWG-based systems operated in the mid-infrared (MIR) range to access stronger absorption lines. However, this typically required the use of MIR PD, which are expensive, bulky, and not easily compatible with fiber-optic systems. In contrast, this PhCW-PTS device utilizes only a near-infrared (NIR) PD, which is mature, cost-effective, and easy integrated with telecom optical fibers. Also, it is possible to extend the system to MIR-PTS by simply replacing the MIR pump source, without using the MIR-PD. The existing NIR PD would still be sufficient for the operation, offering a significant advantage in terms of both cost and integration.

For the fabricated on-chip sensor, we incorporate the PhCW with a high-contrast MZI for a compact and scalable platform for portable gas sensing applications, which has not been achieved in traditional free-space/fiber-based sensors. Beyond gas sensing, the dual-slow-light strategy can be generalized to other dual-light photothermal and photonic platforms, from liquid-phase analyte detection to mid-IR integrated sensors, opening avenues for compact devices in smart-home monitoring and wearable technology.

**Table S3** Comparison of the dual slow-light PhCW-PTS sensor with PhCW/SWG gas sensors

| Year &ref. | Method | Gas λ(nm) | Waveguide type | Γ (%) | $n_g$ | Loss (dB/cm) | L (cm) | MDL @time | NEA (cm$^{-1}$) | NEA·L | Dynamic range |
|---|---|---|---|---|---|---|---|---|---|---|---|
| 2011[28] | DAS | $CH_4$ 1665 | Si-slot PhCW | NA | 30 | NA | 0.03 | 100 ppm | 3.7×10$^{-5}$ | 1.1×10$^{-6}$ | >7 |
| 2021[29] | DAS | $C_7H_8$ 6650 | Si-SWG | 24.3 | 3.83 | 4.3 | 2.84 | 75 ppm | 3.4×10$^{-4}$ | 9.6×10$^{-4}$ | >16 |
| 2022[30] | DAS | $C_3H_6O$ 7330 | Si-SWG | 113 | 4 | 4.7 | 1 | 2.5 ppm (20s) | 2.5×10$^{-5}$ | 2.5×10$^{-5}$ | >1.8×10$^3$ |
| 2024[31] | DAS | $CO_2$ 4345 | SiN-PhCW | 102 | NA | 3.43 | 2 | 1 ppm $^{12}CO_2$ (30s) 20 ppb $^{13}CO_2$ (60s) | 3.4×10$^{-6}$ | 6.8×10$^{-6}$ | 5×10$^4$ |
| 2024[32] | IMS | $C_3H_6O$ 3750 | Si-SWG | 86 | NA | 2.5 | 1 | 50.5 ppm (281s) | 2.8×10$^{-5}$ | 2.8×10$^{-5}$ | >4.9×10$^2$ |
| 2021[33] | IMS | $C_2H_6O$ 3400 | Si-2D PhCW | 2.5 | 73 | NA | 1 | 250 ppb | 1.6×10$^{-6}$ | 1.6×10$^{-6}$ | >4 |
| 2023[15] | IMS | $C_2H_2$ 1531 | Si-2D PhCW$^\#$ Si-1D PhCW | 684 288 | 114 18 | 44 10 | 0.1 | 277 ppm 706 ppm | 3.1×10$^{-4}$ 8×10$^{-4}$ | 3.1×10$^{-5}$ 8×10$^{-5}$ | 7.2×10$^2$ 2.8×10$^2$ |
| 2024[34] | IMS | $CH_4$ 1650 | Si-1D-PhCW | 300 | 4-25 | 20 | 0.1 | 0.055% (17.6s) | 2.4×10$^{-4}$ | 2.4×10$^{-5}$ | 7.3×10$^2$ |
| 2024[35] | IMS | $CO_2$ 4230 | Ge-PhCW | 498 | <50 | 28.91 | 0.08 | 7.56 ppm (18s) | 2.8×10$^{-3}$ | 2.2×10$^{-4}$ | >66 |
| This work | PTS | $C_2H_2$ 1531 | Si-2D PhCW | 234 | 18/81 | 50 | 0.1 | 12 ppm (200s) | 1.4×10$^{-5}$ | 1.4×10$^{-6}$ | >2.5×10$^4$ |

IMS: Intensity modulation spectroscopy. Single slow-light effect is used in other PhCW sensors. SWG: Sub-wavelength grating waveguide, SiN: Silicon nitride, Ge: Germanium, $C_7H_8$: Toluene, $C_3H_6O$: Acetone, $C_2H_6O$: Alcohol. $^\#$In our previous work, the non-suspended single slow-light PhCW is used for pump beam for DAS measurement. In this suspended PhCW, dual slow-light effect is used for both pump and probe beam for PTS measurement.